\documentclass[prd,aps,twocolumn,a4paper,showkeys,nofootinbib]{revtex4-1}

\usepackage{amsmath}
\usepackage{amsfonts}
\usepackage{amssymb}	
\usepackage{graphicx}
\usepackage{bm}
\usepackage{color}
\usepackage{enumitem}
\usepackage{hyperref}
\usepackage{ulem}

\def\non{\nonumber}

\def\GMc2{G M_{\odot} c^{-2}}

\def\lm{{\ell m}}

\def\lm{{\ell m}}

\def\non{\nonumber}

\def\lm{{\ell m}}

\def\TEOBResumS{\texttt{TEOBResumS}}
\def\TEOB{\TEOBResumS}

\def\SEOBNRvq{{\texttt{SEOBNRv4}}}
\def\SEOBNRvqHM{{\texttt{SEOBNRv4\_HM}}}

\def\SEOB{\SEOBNRvq}

\newcommand{\be}{\begin{equation}}  
\newcommand{\ee}{\end{equation}}
\newcommand{\bea}{\begin{eqnarray}}           
\newcommand{\eea}{\end{eqnarray}} 
\newcommand{\beqn}{\begin{eqnarray*}}
\newcommand{\eeqn}{\end{eqnarray*}}
\begin{document}

\title{{\tt TEOBResumS}: assessment of consistent next-to-quasicircular
  corrections and post-adiabatic approximation in multipolar binary black holes waveforms}

\author{Gunnar \surname{Riemenschneider}${}^{1,2}$}
\author{Piero \surname{Rettegno}${}^{1,2}$}
\author{Matteo \surname{Breschi}${}^{3}$}
\author{Angelica \surname{Albertini}${}^{2}$}
\author{Rossella \surname{Gamba}${}^{3}$}
\author{Sebastiano \surname{Bernuzzi}${}^3$}
\author{Alessandro \surname{Nagar}${}^{1,4}$}
\affiliation{${}^1$INFN Sezione di Torino, Via P. Giuria 1, 10125 Torino, Italy}
\affiliation{${}^{2}$ Dipartimento di Fisica, Universit\`a di Torino, via P. Giuria 1, 10125 Torino, Italy}
\affiliation{${}^{3}$Theoretisch-Physikalisches Institut, Friedrich-Schiller-Universit{\"a}t Jena, 07743, Jena, Germany}
\affiliation{${}^4$Institut des Hautes Etudes Scientifiques, 91440 Bures-sur-Yvette, France}

\begin{abstract}

  The use of effective-one-body (EOB) waveforms for black hole
  binaries analysis in gravitational-wave astronomy requires
  faithful models and fast generation times.
  A key aspect to achieve faithfulness is the inclusion of 
  numerical-relativity (NR) informed next-to-quasicircular corrections
  (NQC), dependent on the radial momentum, to the waveform
  and radiation reaction.
  A robust method to speed up the waveform generation is the
  post-adiabatic iteration to approximate the solution of the
  EOB Hamiltonian equations. 
  In this work, we assess the performances of a fast NQC prescription
  in combination to the post-adiabatic method for generating
  multipolar gravitational waves.
  The outlined approach allows a consistent treatment of NQC in
  both the waveform and the radiation-reaction, does not
  require iterative procedures to achieve high faithfulness, 
  and can be efficiently employed for parameter estimation. 
  Comparing to 611 NR simulations, for total mass $10M_\odot\leq M \leq 200M_\odot$ and
  using the Advanced LIGO noise, the model has EOB/NR unfaithfulness well 
  below $0.01$, with 78.5\% of the cases below $0.001$. 
  We apply the model to the parameter estimation of GW150914 
  exploring the impact of the new NQC and of the higher modes up to
  $\ell=m=8$.

\end{abstract}

\maketitle

\section{Introduction}

The continuously increasing sensitivity of gravitational-wave (GW)
detectors~\cite{TheVirgo:2014hva,TheLIGOScientific:2014jea} and the associated compact 
binaries detections~\cite{Abbott:2020niy}
motivate work towards physically complete, precise and efficient
gravitational-wave models.
The effective-one-body (EOB) framework~\cite{Buonanno:1998gg,Buonanno:2000ef,Damour:2000we,Damour:2001tu,Damour:2015isa} 
is a possible approach to the general-relativistic two-body problem that, by construction,
allows the inclusion of perturbative (post-Newtonian, black hole
perturbations) and full numerical relativity (NR) results. It
currently represents a state-of-art approach for modeling waveforms
from binary black holes, conceptually designed to describe the entire
inspiral-merger-ringdown phenomenology of quasicircular binaries~\cite{Nagar:2018gnk,Nagar:2018zoe,Cotesta:2018fcv,Nagar:2019wds,Nagar:2020pcj,Ossokine:2020kjp,Schmidt:2020yuu}
or even eccentric inspirals~\cite{Nagar:2021gss} and hyperbolic captures~\cite{Nagar:2020xsk,Nagar:2021gss}.
In the low-frequency inspiral regime, where NR simulations are not available,
EOB it is the only alternative to improve standard and badly convergent 
post-Newtonian (PN) models for exploring systematics effects in the 
modeling of the radiation reaction~\cite{Gamba:2020wgg}. In the high-frequency 
merger regime, EOB can generate highly faithful waveforms for GW astronomy 
thank to the inclusion of NR information \cite{Nagar:2019wds,Nagar:2020pcj}. 
This paper focuses on a key aspect for EOB models: the consistent
and efficient inclusion of NR information in the multipolar waveform.

Current EOB models are informed by NR in two separate ways: (i) on the one hand,
through {\it EOB flexibility parameters}~\cite{Damour:2002qh} that allow to improve the
conservative part of the dynamics, i.e. typically as effective high-order terms in the 
orbital, spin-orbit or spin-spin part sector of the EOB Hamiltonian; (ii) on the other hand, 
through next-to-quasi-circular (NQC) corrections to the multipolar waveform 
(and flux)~\cite{Damour:2007xr,Damour:2007vq,Damour:2008te,Damour:2009kr}.
The latter enter as multiplicative factors, that depend on the radial
motion, and correct the EOB factorized quasicircular waveform~\cite{Damour:2008gu,Messina:2018ghh}
multipole by multipole, so to introduce effective, NR-tuned, modifications to both the amplitude and the phase. 
NQC corrections are essential to improve the analytical quasicircular waveform during 
the late plunge up to merger; they also guarantee a smooth transition to the subsequent 
ringdown phase. Importantly, NQC parameters are the largest set of data inferred from NR. 
For example, the spin-aligned \TEOB{} model uses NR information to determine 2 parameters 
(one orbital and one spin-orbital)\cite{Nagar:2020pcj} for the spin-aligned effective 5PN Hamiltonian, 
but  36 parameters (two for amplitude and two for phase) for the NQC-corrected multipolar waveform, 
that can have up to {\it 9 multipoles}\footnote{This procedure is robust as long as spins are mild, 
say up to $\sim 0.5$. In the nonspinning case it is even possible to
complete through merger and ringdown a typically negligible mode as the $(4,1)$. For large spins,
some modes like $(2,1)$, $(4,3)$ or $(4,2)$ may be inaccurate due to the delicate interplay between the 
strong-field dynamics and the NQC factor.} completed through merger 
and ringdown~\cite{Nagar:2020pcj}: $(\ell,|m|)=\{(2,2),(2,1),(3,3),(3,2),(3,1),(4,4),(4,3),(4,2),(5,5)\}$.
All higher modes up to $\ell=8$ can also be optionally generated by the model, although currently without
the NR-informed merger ringdown~\cite{Nagar:2019wds}.
In the spin-aligned \SEOBNRvq{}~\cite{Bohe:2016gbl} and \SEOBNRvqHM{}~\cite{Cotesta:2018fcv} 
the amount of information inferred from NR is similar, although it is
included differently. 
In particular: (i) there are 3 flexibility parameters entering the Hamiltonian~\cite{Bohe:2016gbl} 
(that is different from the \TEOBResumS{} one~\cite{Rettegno:2019tzh}); 
(ii) for each waveform multipole there are 5 NQC parameters (3 for the amplitude and 2 for the phase\footnote{This is because \SEOBNRvq{} also imposes that the EOB and NR 
amplitude {\it curvature} coincide near merger.}), 
for a total of 25 parameters since the modes completed through merger 
and ringdown are $(\ell,|m|)=\{(2,2),(2,1),(3,3),(4,4),(5,5)\}$. In addition, \SEOBNRvqHM{} needs 
two more effective corrections to the $(2,1)$ and $(5,5)$ amplitudes that are calibrated to NR.

To achieve internal consistency between the waveform and the radiation reaction 
in the EOB equations of motion, the NQC amplitude factor should be also 
incorporated within the radiation reaction force, i.e. the flux of mechanical angular momentum. 
A possible approach to this problem is to iterate the dynamics several time, 
updating the values of NQC parameters at each step, until their values are seen to 
converge~\cite{Damour:2009kr,Damour:2012ky}. This procedure, though 
necessary from the physical point of view, cannot be part of a waveform 
generator for parameter estimation, as it would increase the global 
computational time at least by a factor four. Yet, it is important because,
as we will see below, it also yields a fractional agreement between the NR and EOB 
angular momentum flux $\lesssim 1\%$ even during the late-inspiral and plunge regime.
One way out is simply to avoid this iterative procedure and keep radiation reaction
without the NQC corrective factor. This route is the one implemented in
\SEOB{}~\cite{Bohe:2016gbl}, but evidently the model lacks of
self consistency between radiation reaction and waveform\footnote{This self-consistency 
problem is even amplified in \SEOBNRvqHM{} because the PN information incorporated
in the higher waveform multipoles is not the same as the one present in radiation reaction. 
Doing so, would have implied a new NR recalibration of the \SEOB{} dynamics, a route not 
followed for \SEOBNRvqHM{}. By contrast, this has always been the case for \TEOB{} due 
to the lighter and simpler NR calibration procedure.}.

Reference~\cite{Nagar:2020pcj} (hereafter Paper~I), shows that the $(2,2)$
mode of \TEOB{} {\it with iterated NQC} corrections achieves an overall EOB/NR
unfaithfulness for total mass $10M_\odot\leq M \leq 200M_\odot$ is always below $0.5\%$,
with one single outlier grazing the $0.85\%$ level. \SEOBNRvq{}, {\it without the iterated NQC} 
at most grazes $1\%$, although it has been tested on only 114 spin-aligned 
NR waveforms~\cite{Bohe:2016gbl} up to $q=10$. This number is six time smaller 
than the testing sample of \TEOBResumS{}, that is also pushed up to mass ratio $q=18$.

In this paper, we describe the NQC fitting procedure used in \TEOB{} in
order to obtain a consistent (waveform and flux) NQC term without the
iteration procedure. This NQC treatment is the default option in the most
recent version of \TEOBResumS{}, that incorporates 
higher modes~\cite{Nagar:2020pcj} and has been already used 
in~\cite{Breschi:2021wzr}, although not reported before. 
For simplicity we will refer to this version as {\tt v2}. By contrast, the {\tt v1} tag refers 
to the first implementation of \TEOBResumS{}~\cite{Nagar:2018zoe}. 
We also present an updated faithfulness assessment
of the \TEOB{} $\ell=m=2$ waveform against a large set of NR simulations
where we include for the first time:
(i) the new NQC fits;
(ii) the (iterated) post-adiabatic approximation to the dynamics~\cite{Damour:2012ky,Nagar:2018gnk,Nagar:2018plt,Rettegno:2019tzh}.

The post-adiabatic (PA) approximation is a robust method to solve the 
EOB Hamiltonian equations by an iterative analytical procedure rather than 
solving numerically the set of ODEs.
The PA was shown to be crucial for parameter estimation with \TEOB{}, 
both for black holes and neutron stars~\cite{Gamba:2020wgg,Gamba:2020ljo,Breschi:2021wzr}. 
In particular, the PA is a simple, flexible and robust alternative to 
surrogate methods~\cite{Lackey:2016krb,Cotesta:2020qhw}. 
By using this approach, the dynamics computation can become up to 20 times faster and 
its employment is among the reasons why the \TEOB{} computational cost is generally one 
order of magnitude smaller than the {\tt SEOBNRv4HM}~\cite{Cotesta:2018fcv} one.
This method is implemented in the most recent stand-alone release of \TEOB{} as well 
as in the {\tt v1} release within the LIGO Algorithm Library (LAL)~\cite{lalsuite}.
We demonstrate the use of the NQC fits and of the PA approximation 
in parameter estimation on GW150914, notably using the multipolar waveform with 
all modes up to $\ell=m=8$. In particular, the possibility of doing PE with and without 
NQC fits allows us to analyze in detail a very specific source of analytical systematics in
waveform modeling.

This paper is organized as follows. Section~\ref{sec:nqc} reviews the motivations and 
structure of the NQC correction and the new fits. 
Sec.~\ref{sec:F} discusses the validation of the production setup of
\TEOB{} with the new NQCs and the PA against 595 SXS and 19 BAM waveforms.
In Sec.~\ref{sec:timing} we give an account of the \TEOB{} waveform generation time.
Finally, Sec.~\ref{sec:gw150914} presents the application to GW150914 analysis.
After the conclusions, the paper has two appendices:
Appendix~\ref{sec:F:iter} reports the unfaithfulness plots of Paper~I to
facilitate the comparison with the new results; 
Appendix~\ref{sec:fit} contains all the details on the new NQC fits.

\section{EOB Next-to-quasicircular corrections} 
\label{sec:nqc}

\begin{figure}[t]
  \begin{center}
    \includegraphics[width=0.48\textwidth]{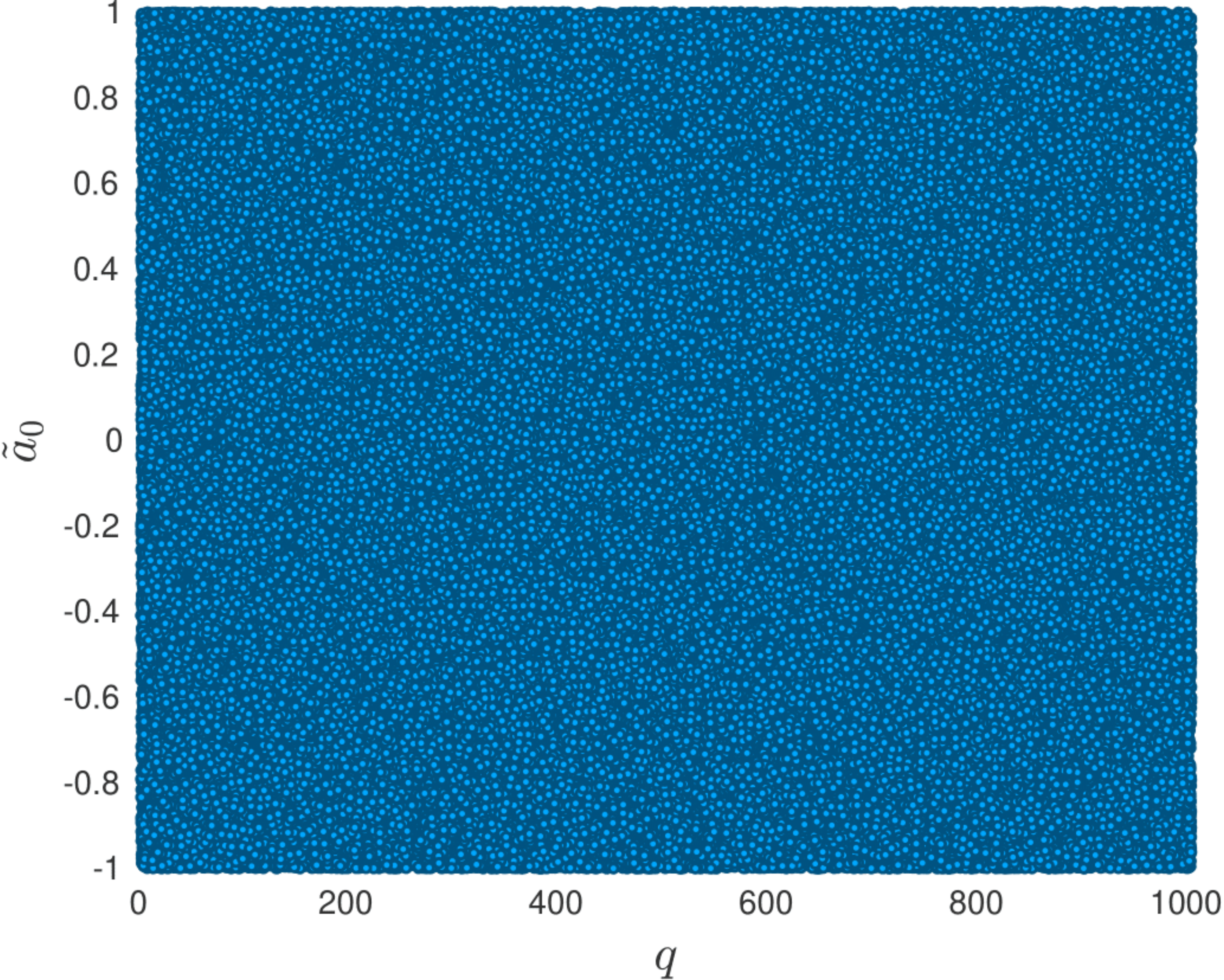}
        \caption{Robustness of the \TEOBResumS{} implementation with
      higher modes, PA dynamics and NQC corrections fits in the flux.
      Waveforms (blue markers) are generated with no failures
      for 250,000 binaries with parameters uniformly sampled at random in the ranges $1\leq q\leq 1000$ and $-1 < \chi_i < 1$. 
      The spin parameter on the $y$-axis is defined as $\tilde{a}_0=\chi_{\rm eff}=(m_1\chi_1 + m_2 \chi_2)/M$.  
      \label{fig:parspace_coverage}}
  \end{center}
\end{figure}

Next-to-quasi-circular corrections were introduced in the first EOB analysis of 
the transition from inspiral to plunge, merger and ringdown in the test-particle 
limit~\cite{Damour:2007xr}. They were originally conceived as an effective noncircular 
correction to the flux of mechanical angular momentum $\cal{F}_\varphi$,
 so to consistently model it during the plunge up to merger 
 (see Fig.~2 in Ref.~\cite{Damour:2007xr}). In subsequent
EOB/NR works~\cite{Damour:2007vq,Damour:2008te} they were moved to the
$(2,2)$ waveform in order to achieve an optimal EOB/NR amplitude and phase agreement
at merger and ease the attachment of the ringdown part. Finally, Ref.~\cite{Damour:2009kr}
introduced the current paradigm, within \TEOBResumS{}, of having them 
in both the $(2,2)$ waveform and radiation reaction, with the iterative procedure to 
consistently determine the effective NQC parameters entering the $(2,2)$ amplitude.
More precisely, each factorized and resummed~\cite{Damour:2008gu} EOB 
waveform mode $(\ell,m)$ is dressed by a multiplicative contribution $\hat{h}^{\rm NQC}_\lm$
as
\be
h_{\ell m} = h_{\ell m}^{(N,\epsilon)}\hat{h}_{\ell m}\hat{h}^{\rm NQC}_{\ell m}\ ,
\ee
where $h_{\ell m}^{(N,\epsilon)}$ is the Newtonian prefactor with
parity $\epsilon$ and $\hat{h}_{\ell m}$ the relativistic
correction. The NQC factor is parametrized by 
four parameters $(a^{\ell m}_1,a^{\ell m}_2, b^{\ell m}_1,b^{\ell
  m}_2)$,
\begin{align}
\hat{h}^{\rm NQC}_{\ell m} &=
\left(1+a_1^{(\ell,m)}n^{(\ell,m)}_1+a_2^{(\ell,m)}n^{(\ell,m)}_2\right)\non\\
&\times e^{{\rm i}(b_1^{(\ell,m)}n^{(\ell,m)}_3+b_2^{(\ell,m)}n^{(\ell,m)}_4 )} \ ,
\end{align}
where $n_i^{(\ell,m)}$ are functions depending on the radial velocity
and acceleration, see Eqs.(3.32)-(3.35) of~\cite{Nagar:2019wds} 
and Ref.~\cite{Damour:2012ky}.
Parameters  $(a^{(\ell,m)}_1,a^{(\ell, m)}_2)$ determine the 
NQC of the amplitude's multipole $(\ell,m)$, while $(b^{(\ell,m)}_1,b^{(\ell,m)}_2)$
determine the NQC to the phase and frequency of the multipole $(\ell,m)$.
The parameters $(a^{(2,2)}_1,a^{(2,2)}_2)$ play a special role as they are
those also included in the radiation reaction~\cite{Damour:2009kr}. 
Their best values are determined by an iterative procedure, e.g. the one of Paper~I. 
The parameters  $(a^{(\ell,m)}_1,a^{(\ell,m)}_2,b^{(\ell,m)}_1,b^{(\ell,m)}_2)$
with $(\ell,m)\neq (2,2)$ are instead best generated by solving a set of four 
coupled algebraic equations and imposing NR-informed fits of amplitude, 
frequency and their first derivatives around 
merger \cite{Damour:2009kr,Damour:2012ky,Nagar:2017jdw,Nagar:2018zoe,Nagar:2019wds,Nagar:2020pcj}.


%

\subsection{Fitting NQC parameters $(a^{(2,2)}_1,a^{(2,2)}_2)$}
\label{sbsec:newfits}
\begin{figure}[t]
  \begin{center}
    \includegraphics[width=0.49\textwidth]{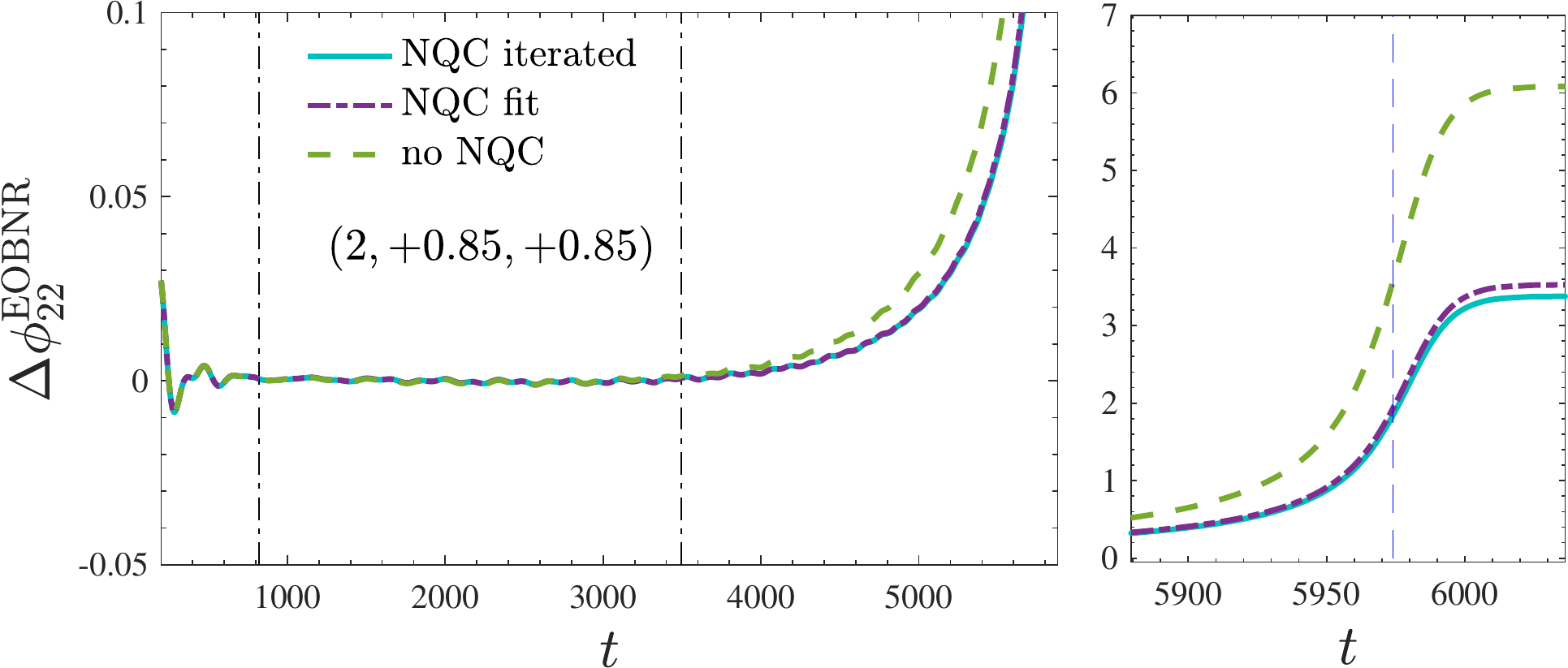}
        \caption{Comparison between EOB/NR phase differences $\Delta\phi_{22}^{\rm EOBNR}\equiv \phi_{22}^{\rm EOB}-\phi_{22}^{\rm NR}$ 
        for dataset SXS:BBH:0257, configuration $(q,\chi_1,\chi_2)=(2,+0.85,+0.85)$. Waves are aligned in the early inspiral, on the time interval 
        indicated by the vertical dash-dotted lines in the left panel. The vertical line in the right panel marks merger time, corresponding to the
        peak of the $\ell=m=2$ amplitude. The curves correspond to: (i) iterated NQC parameters $(a_1^{(2,2)},a_2^{(2,2)})$ in radiation reaction 
        as in Paper~I; fitted NQC parameters;  no NQC parameters in radiation reaction. The corresponding maximum values of the EOB/NR 
        unfaithfulness $\bar{F}$ from Eq.~\eqref{eq:barF} are $0.414\%$, $0.456\%$ and $1.7\%$ respectively. See the corresponding 
        $\bar{F}(M)$ curves in Figs.~\ref{fig:barF} and \ref{fig:barF_iter} below.                
      \label{fig:nqc_test}}
  \end{center}
\end{figure}
\begin{figure}[t]
  \begin{center}
    \includegraphics[width=0.49\textwidth]{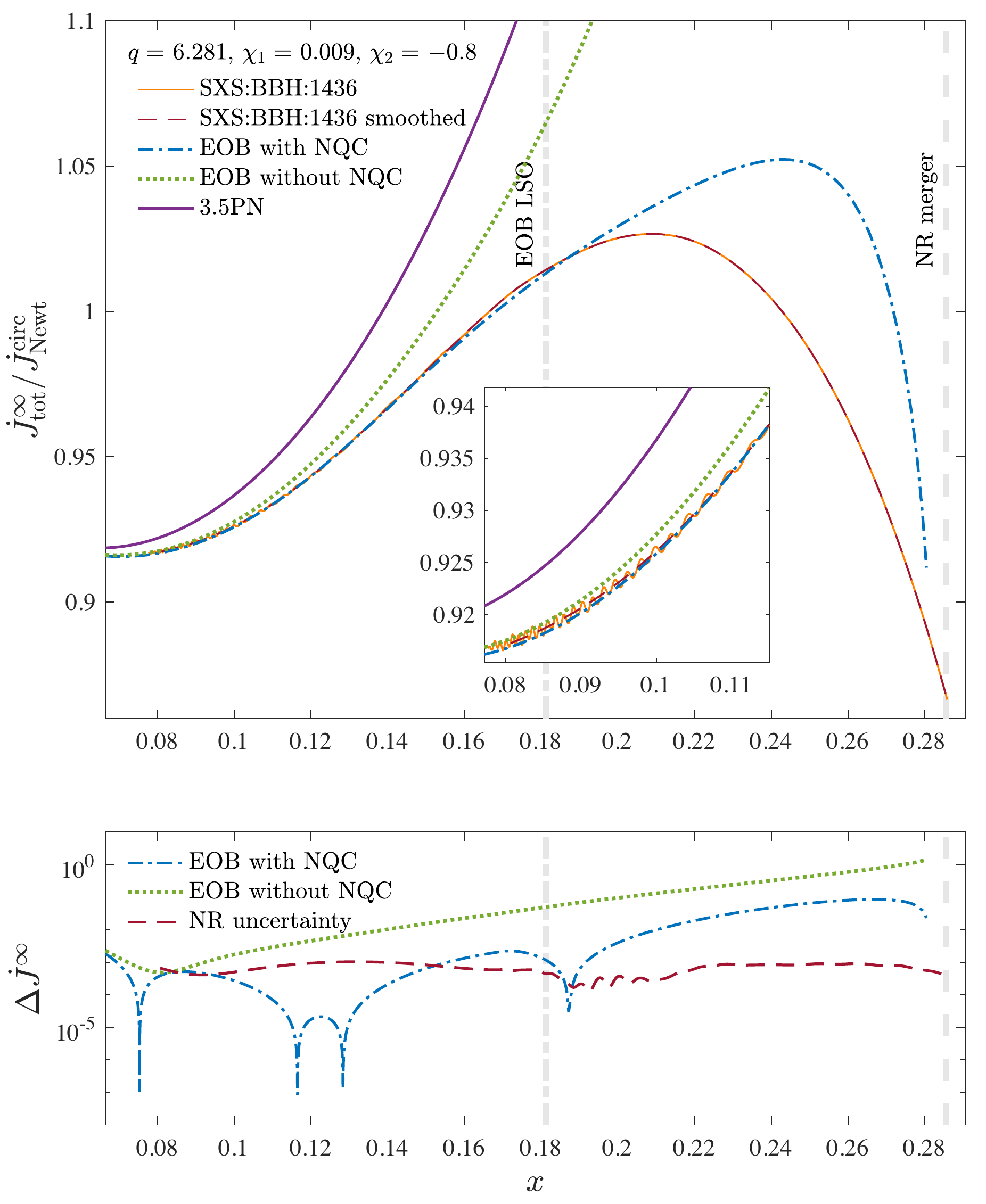}
        \caption{EOB/NR comparison of the fluxes of angular momentum at infinity for a  demonstrative configuration 
        $(q,\chi_1,\chi_2)=(6.281,0.009,-0.8)$ corresponding to SXS:BBH:1436 dataset. Top panel: Newton-normalized
        fluxes versus PN frequency parameter $x$. Bottom panel: EOB/NR fractional differences. The (iterated) EOB 
        flux with the NQC correction factor shows  $\sim 10^{-3}$ fractional agreement  with the NR flux
        up to the  EOB last stable orbit. A much larger difference is found in the absence of NQC correction factor.
        The Taylor-expanded 3.5PN flux is also shown to highlight the power of the EOB resummation procedures. 
        See text for details. \label{fig:flux_comparison}}
  \end{center}
\end{figure}

The high NR-faithfulness of \TEOBResumS{} in Paper~I depends on the EOB flexibility
functions $(a_6^c,c_3)$ that are NR-informed under the conditions that $(a^{(2,2)}_1,a^{(2,2)}_2)$ 
are determined from the iterative procedure. Dropping this would imply a worsening 
of the global EOB/NR agreement (see below). As a consequence, we need to
construct accurate fits of $(a^{(2,2)}_1,a^{(2,2)}_2)$  all over the parameter space
so to obtain EOB/NR unfaithfulness similar to the iterative procedure while
not requiring iterations.
To do so, we proceed as follows. First, the parameters $(a^{(2,2)}_1,a^{(2,2)}_2)$ 
are determined with the same iterative procedure of Paper~I for $2291$ simulations
up to mass-ratio of $q=30$ with aligned spins up to $\chi_1=\chi_2=\pm 0.99$.
Second, the values $(a^{(2,2)}_1,a^{(2,2)}_2)$ are fitted across the
parameter space. The latter is divided in four different regions:
\begin{enumerate}
\item[(i)] Nonspinning sector, $\chi_1=\chi_2=0$ 
\item[(ii)] Spinning sector, equal-mass sector with $\nu>0.2485$ 
\item[(iii)] Spinning sector, $0.16\leq\nu<0.2485$ 
\item[(iv)] Spinning sector, with $\nu\leq 0.16$.
\end{enumerate}
In each region different templates are employed to better capture the
functional behavior of $(a^{(2,2)}_1,a^{(2,2)}_2)$. All fits are done 
using as single spin parameter the standard spin combination
\be
\hat{S} \equiv \dfrac{S_1+S_2}{M^2}=X_1^2 \chi_1 + X_2^2 \chi_2 \ ,
\ee
where $S_i$ are the dimensionful individual spins, $\chi_i \equiv S_i/m_i^2$ 
are the dimensionless spins and $X_i\equiv m_i/M$. The spin parameter $\hat{S}$
is actually used in the fits only for the equal-mass case. In the other situations,
it looks more flexible to incorporate some $\nu$-dependence and use 
instead\footnote{This variable, called $\chi$, is used in various fits of merger and postmerger quantities 
entering the \SEOBNRvq{} model~\cite{Bohe:2016gbl}.}
\begin{align}
\hat{S}_\nu\equiv \dfrac{\hat{S}}{1-2\nu} \  .
\end{align}
All the details of the fitting procedure are in Appendix~\ref{sec:fit}.

Our NQC implementation has been extensively tested to check its robustness all over the 
parameter space. Fig.~\ref{fig:parspace_coverage} illustrates that the
new NQC implementation never failed for 420,000 binary configurations
drawn from random distributions of spins $-1<\tilde{a}_0<+1$ and mass
ratios $1\leq q\leq 1000$. The EOB runs in the figure are generated with the 
PA method, computing the dynamics up to the dimensionless radius 
$r = R/GM = 14$ on a grid with $dr = 0.1$ using the 8th PA 
order~\cite{Nagar:2018gnk}. The other NR-informed EOB parameters are the 
same as in~\cite{Nagar:2020pcj} and corresponds to the default configuration 
of \TEOB{} for parameter estimation.

\begin{figure*}[t]
  \begin{center}
    \includegraphics[width=0.24\textwidth]{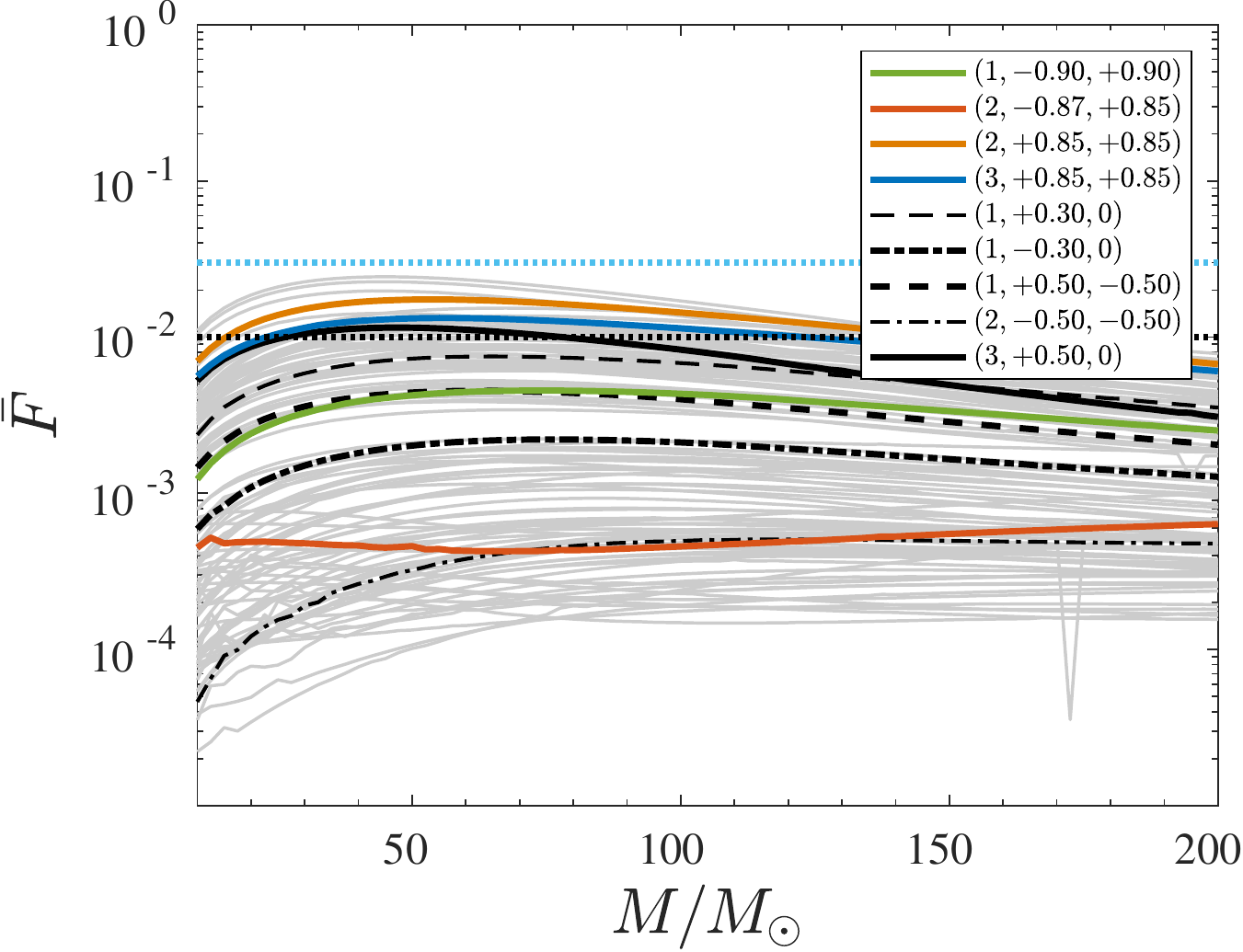}
    \includegraphics[width=0.24\textwidth]{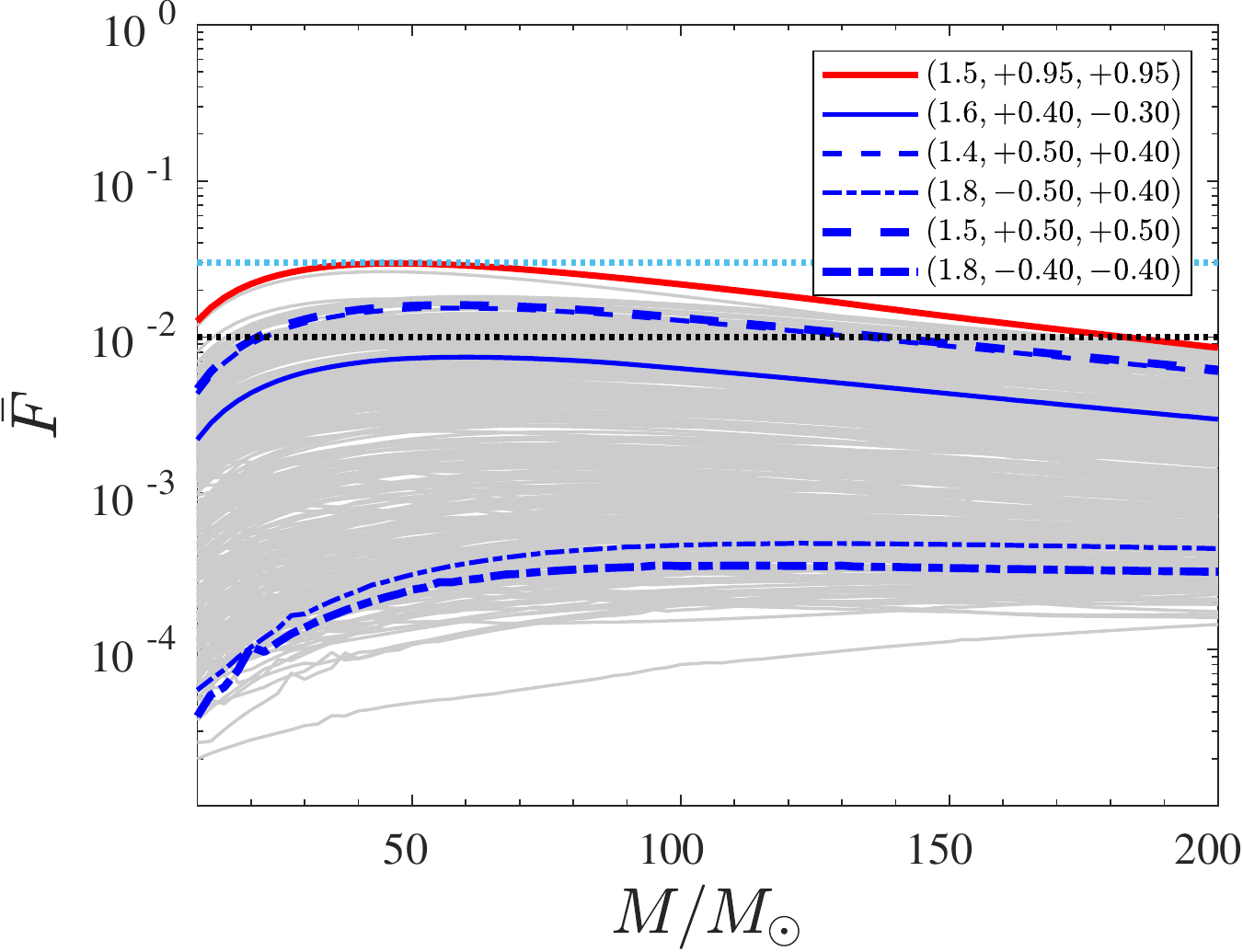}
    \includegraphics[width=0.24\textwidth]{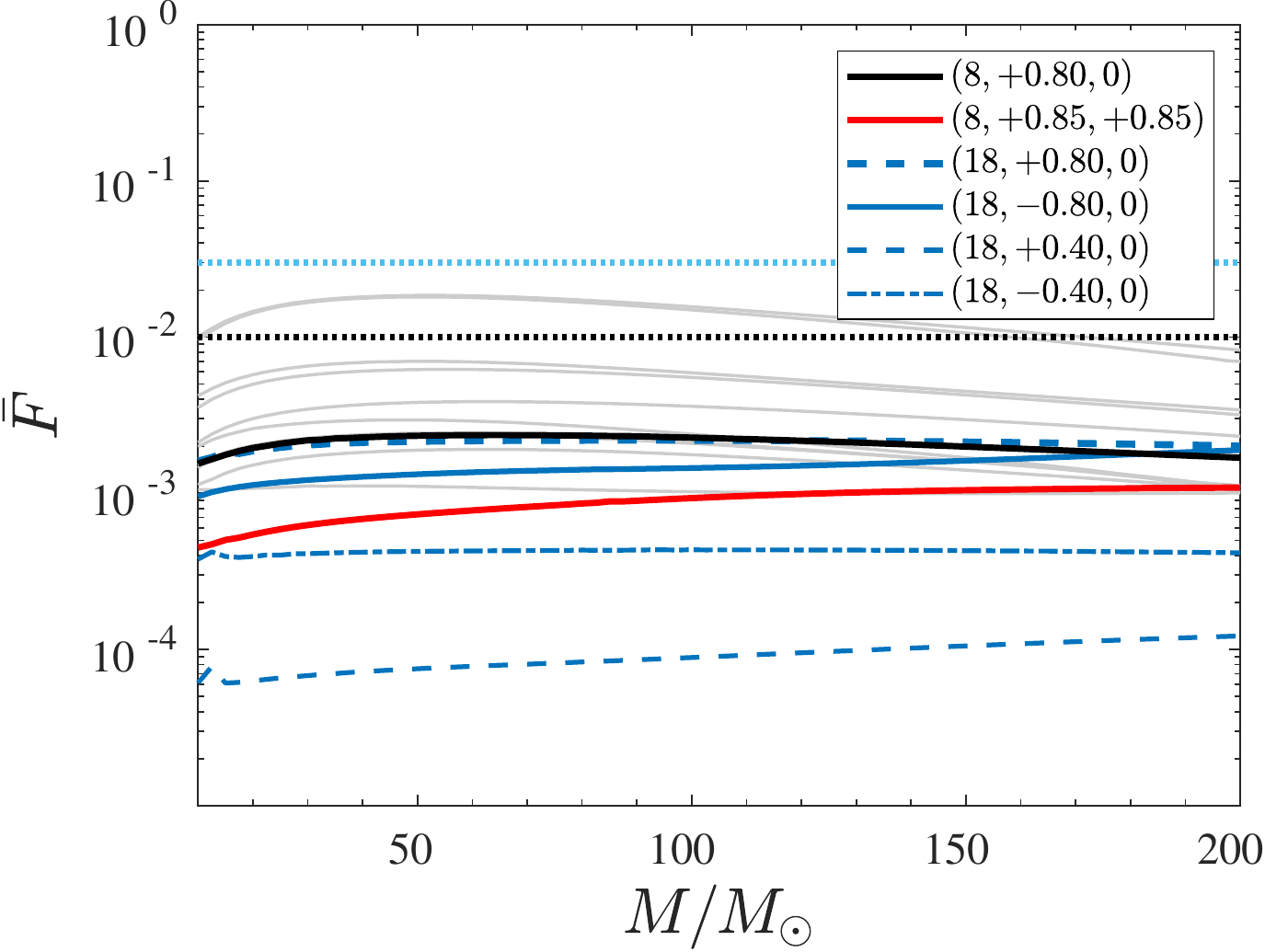} 
    \includegraphics[width=0.24\textwidth]{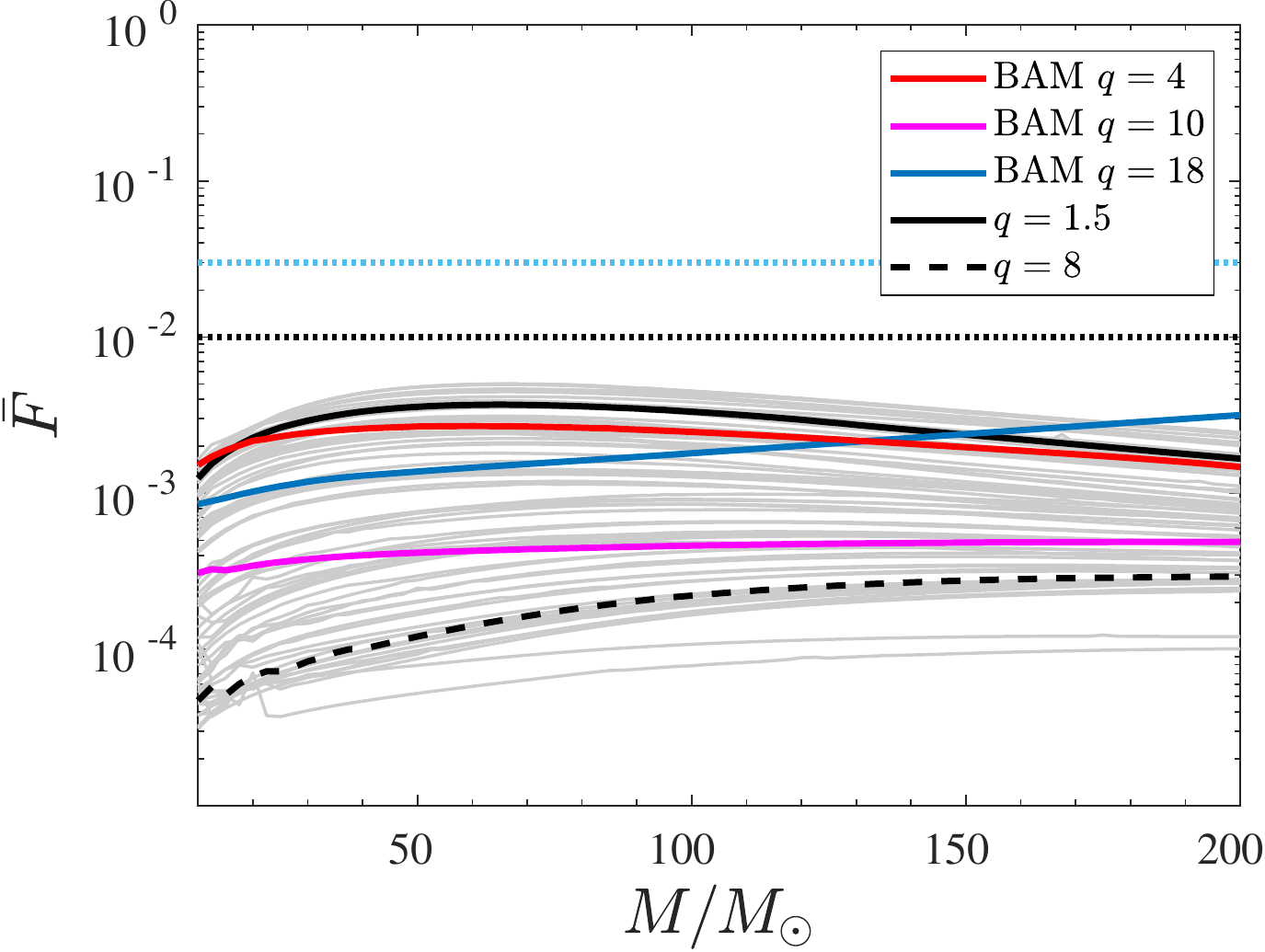} \\
     \includegraphics[width=0.24\textwidth]{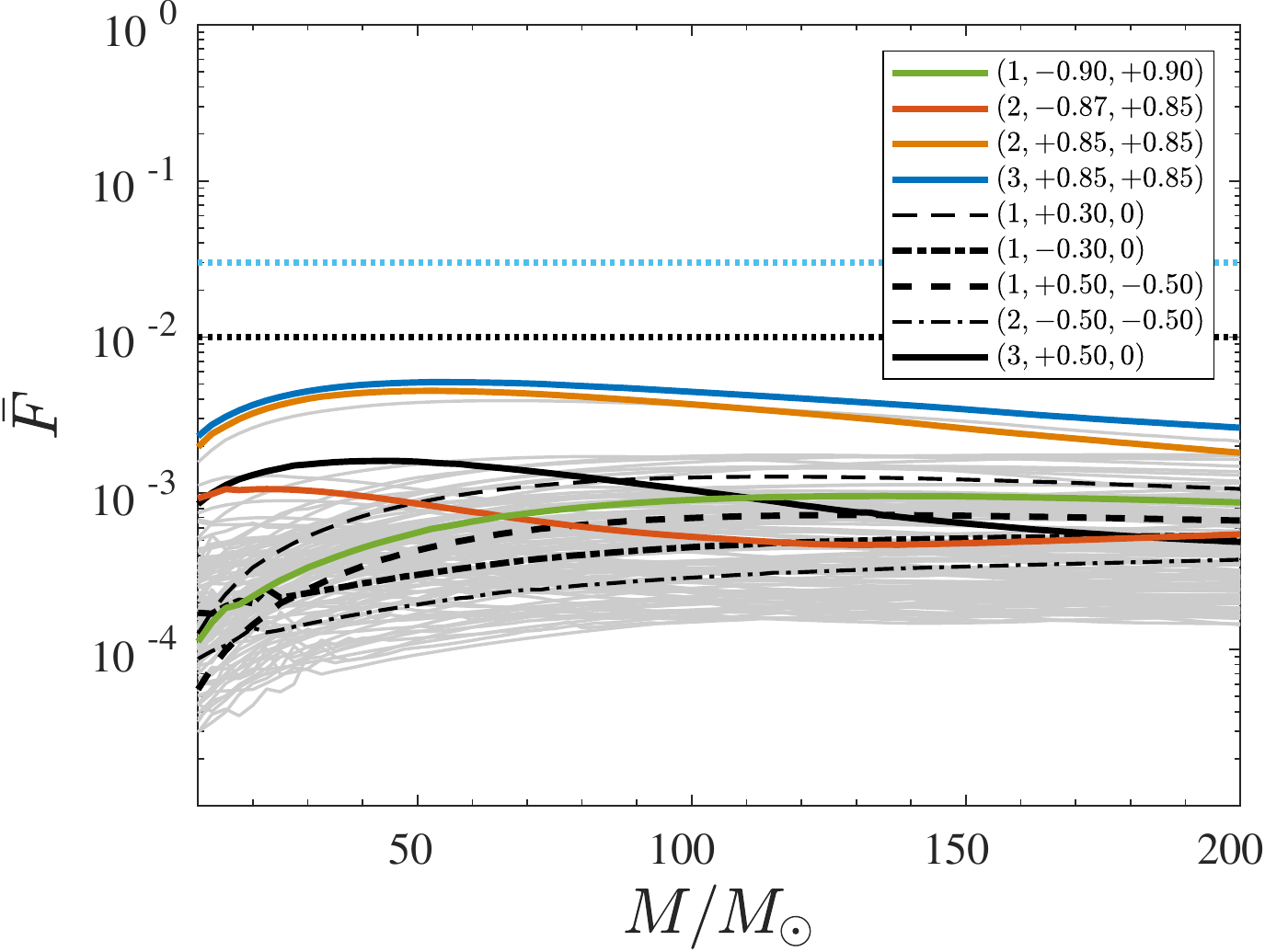}
    \includegraphics[width=0.24\textwidth]{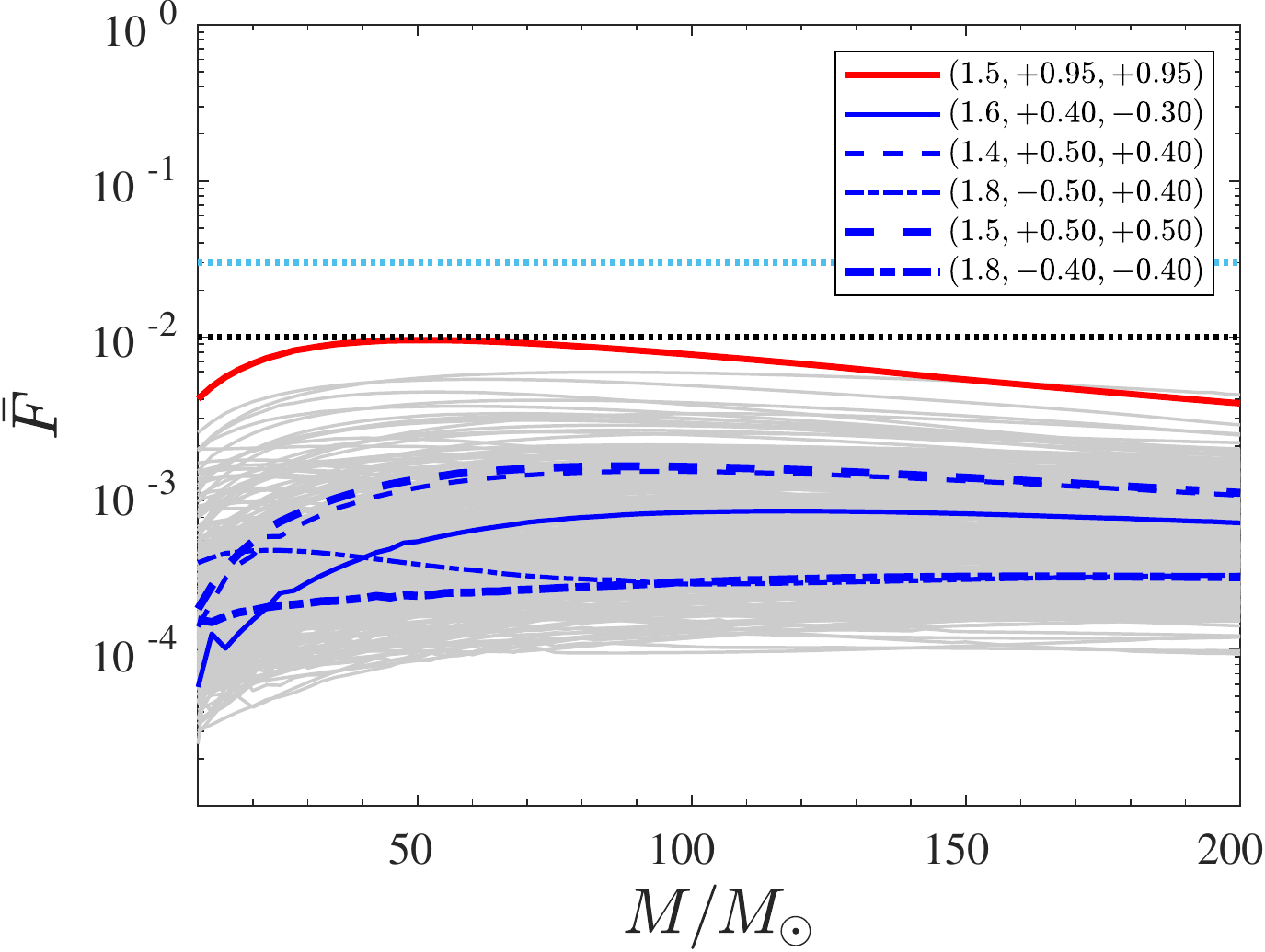}
    \includegraphics[width=0.24\textwidth]{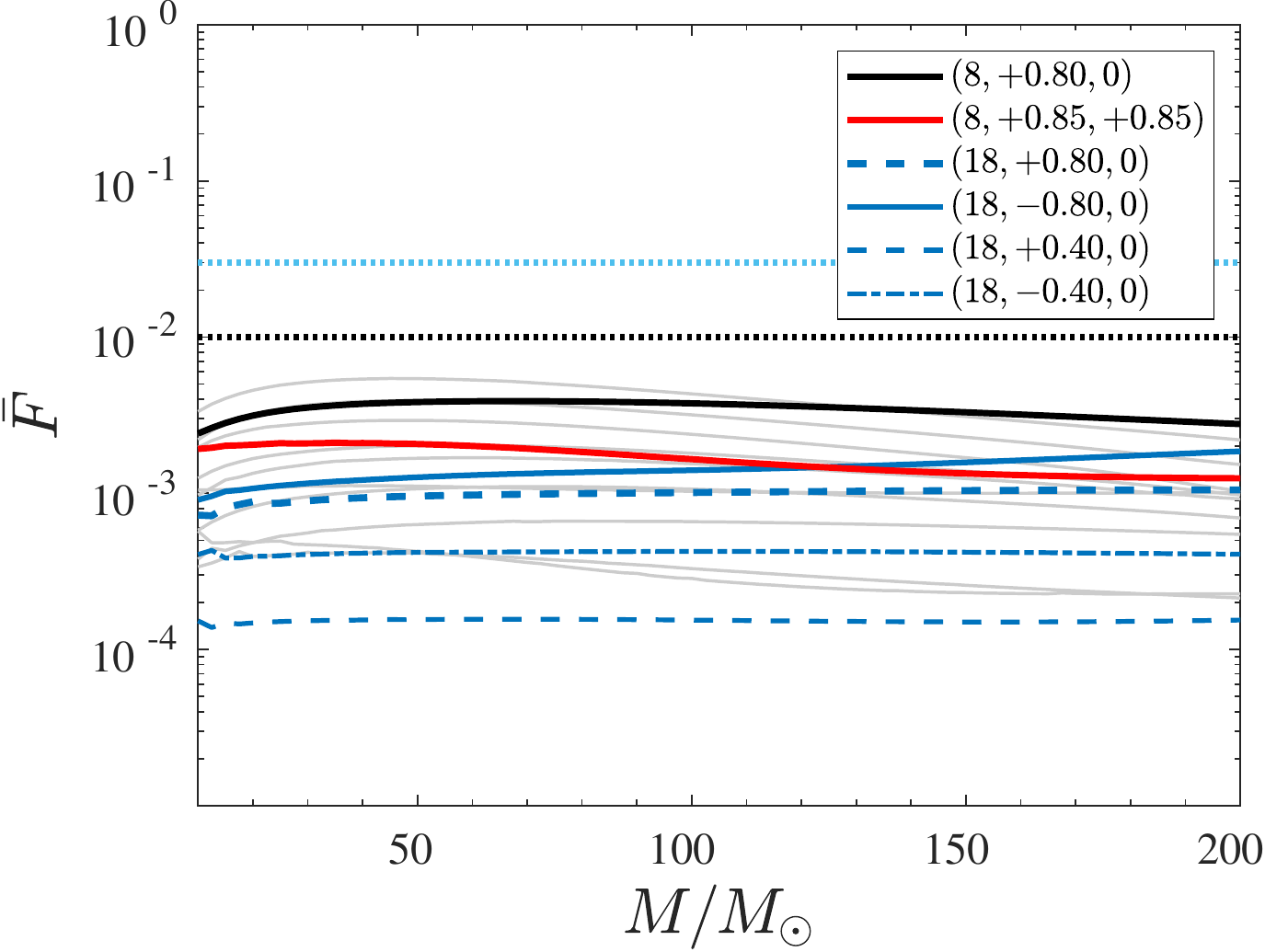}
    \includegraphics[width=0.24\textwidth]{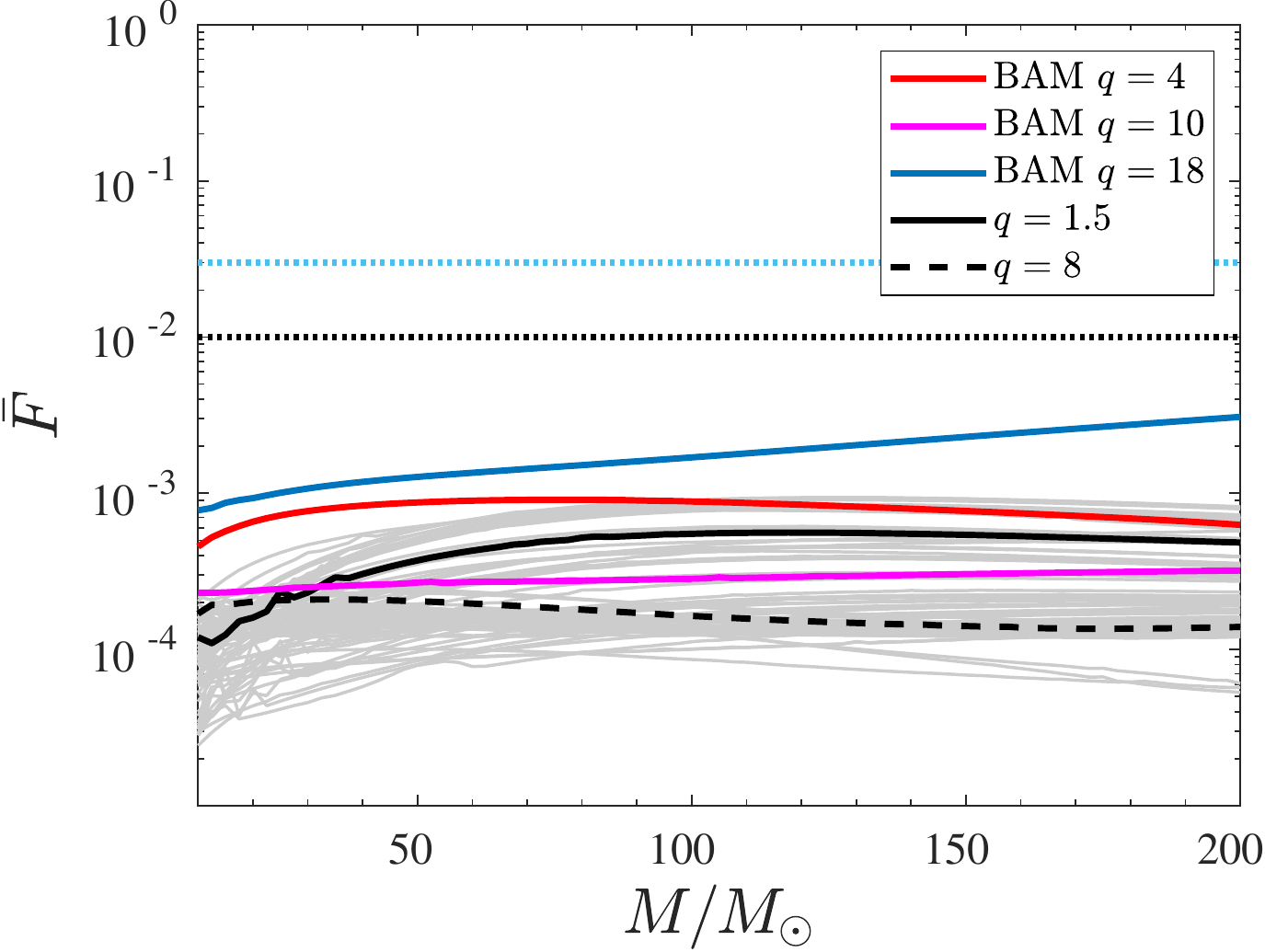}
    \caption{EOB/NR unfaithfulness for the $\ell=m=2$ mode using all currently available spin-aligned 
                  SXS NR simulations and a bunch of BAM simulations. Top row: \TEOBResumS{}
                  {\it with} the PA approximation for the inspiral and {\it without} NQC corrections in radiation reaction.
                  Bottom row: \TEOBResumS{} {\it with} the PA approximation for the inspiral and {\it with} the NQC 
                  parameters obtained by the fit in radiation reaction. From left to right, the columns use the following
                  NR data: SXS spin-aligned waveforms publicly released before February 3, 2019;
                  SXS spin-aligned waveforms publicly released after February 3, 2019; spin-aligned BAM data; nonspinning configurations.
                  The quality of the EOB performance with the NQC fits is very good and essentially equivalent to the outcome of the 
                  exact iterative procedure of Ref.~\cite{Nagar:2020pcj}, that is reported in Fig.~\ref{fig:barF_iter} in Appendix for completeness.}                                    
    \label{fig:barF}
  \end{center}
\end{figure*}

\begin{figure*}[t]
  \begin{center}
    \includegraphics[width=0.4\textwidth]{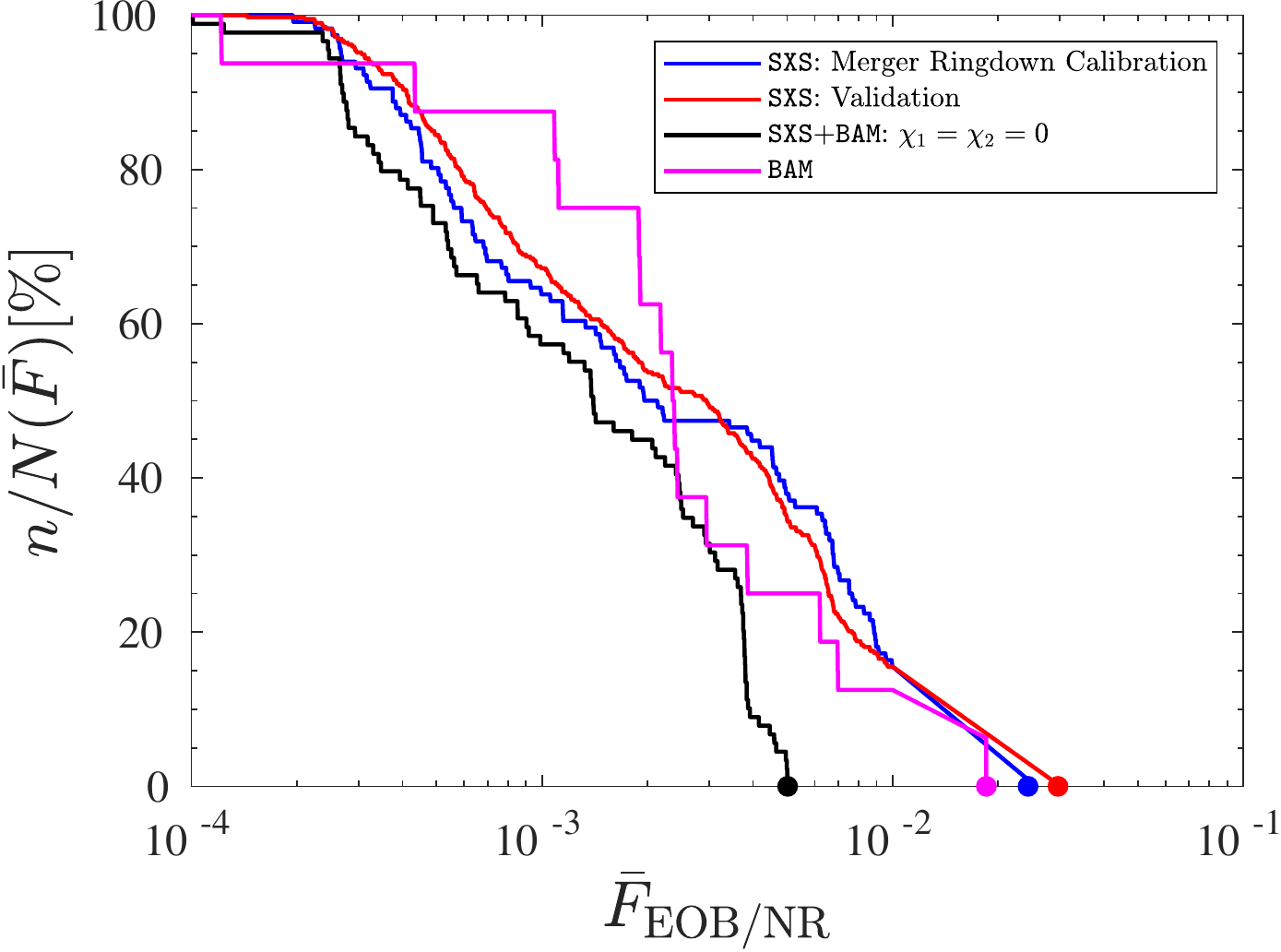}
    \hspace{10mm}		
    \includegraphics[width=0.4\textwidth]{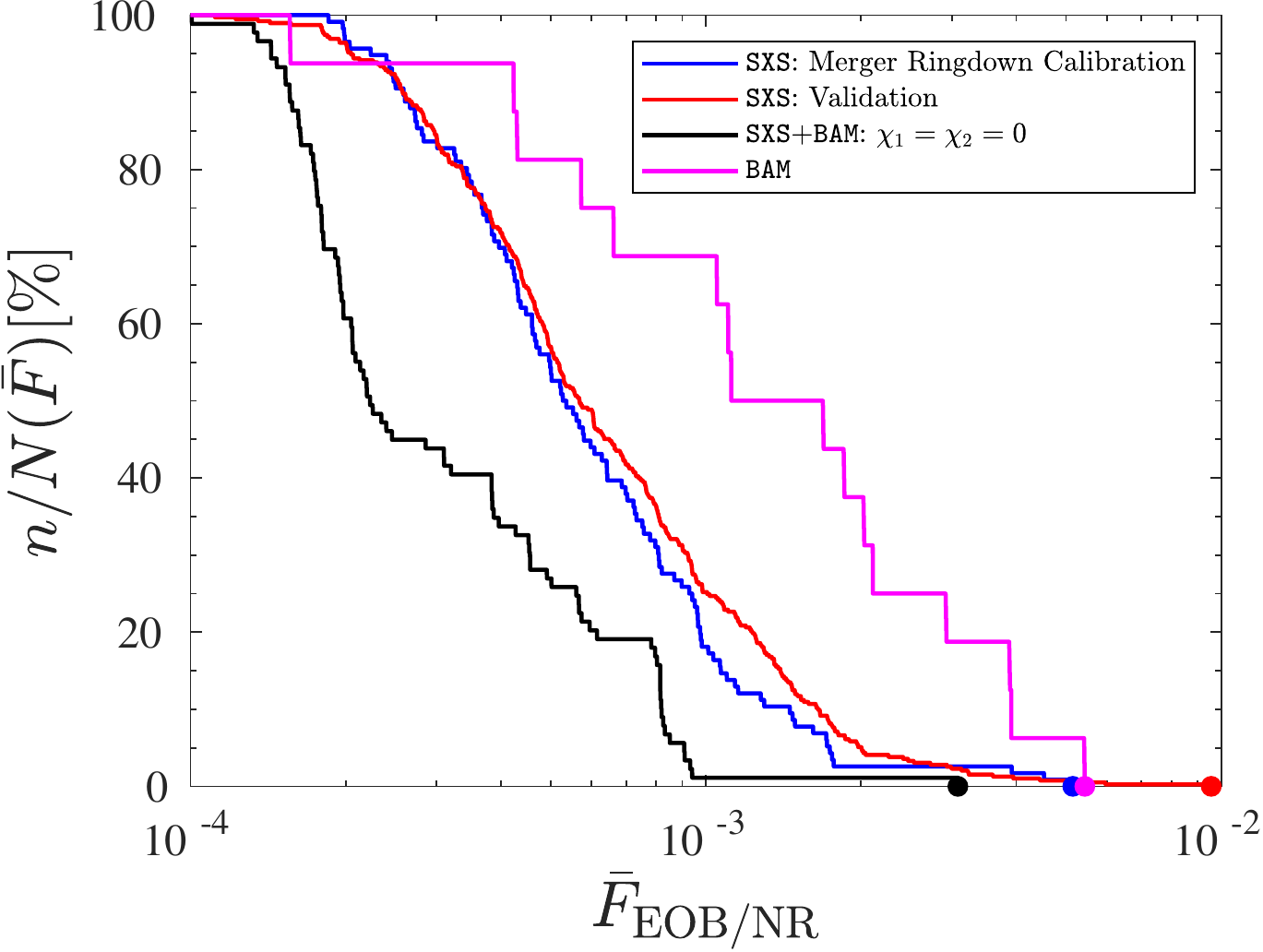}
    \caption{\label{fig:histo} Summary histogram of EOB/NR
      unfaithfulness $\bar{F}_{\rm EOB/NR}$ over the full NR database 
      of 611 simulations, without NQC fits (top panel) and with fits (bottom panel). 
      The various SXS subsets, nonspinning (black online, 89 waveforms), 
      merger-ringdown calibration 
    (blue online, 116 spin-aligned waveforms)  and validation 
    (red online, 388 spin-aligned waveform) as defined in Paper~I are presented
    separately. The plot shows  the fraction (expressed in $\%$) 
    $n/N_{\rm set}$, where $N_{\rm set}$ is the total number of waveforms in a given 
   NR-waveform set and $n$ is the number of waveforms, in the same set, that, 
   given a value $\bar{F}$, have  $\bar{F}^{\rm max}_{\rm EOB/NR}\geq \bar{F}$. 
   The colored marker highlight the largest values in each NR dataset.}
  \end{center}
\end{figure*}

\begin{table*}[t]
	\caption{\label{postresults} GW190514 analysis and main parameters intervals. 
		We report the median and $90\%$ credible region for the parameters extracted from the  posterior distribution. 
		Explicitly, the total mass $M$, the chirp mass $\mathcal{M_{\rm c}}$, the individual masses $m_i$, the mass ratio $q$, the dimensionless 
		spins $\chi_i\equiv S_i/m_i^2$ and their combination $\chi_{\rm eff} = \tilde{a}_0=(m_1 \chi_1 + m_2 \chi_2)/M$, the luminosity distance $D_{\rm L}$, 
		the inclination angle $\iota$, the right ascension $\alpha$ and declination $\delta$. In the last row, we show the logarithmic Bayes' factor 
		with its standard deviation.}
	\begin{center}
		\begin{ruledtabular}
			\begin{tabular}{ccccccc}
				&     22+NQCfit     &  22+noNQCfit   &    LM+NQCfit   & LM+noNQCfit    &    HM+NQCfit & HM+noNQCfit \\
				\hline
				$M~{\rm [M_\odot]}$ & $72.11_{-2.55}^{+2.79}$ & $73.42_{-2.67}^{+2.88}$ & $72.48_{-2.31}^{+3.51}$ & $72.87_{-2.85}^{+3.66}$ & $73.30_{-2.74}^{+3.18}$ & $72.86_{-2.88}^{+3.31}$ \\
				$\mathcal{M}_{\rm c}~{\rm [M_\odot]}$ & $31.16_{-1.20}^{+1.25}$ & $31.72_{-1.31}^{+1.30}$ & $31.39_{-1.12}^{+1.50}$ & $31.54_{-1.26}^{+1.59}$ & $31.75_{-1.26}^{+1.42}$ & $31.54_{-1.25}^{+1.46}$ \\
				$m_1~{\rm [M_\odot]}$ & $39.67_{-3.32}^{+4.31}$ & $40.06_{-3.07}^{+3.79}$ & $38.83_{-2.35}^{+3.73}$ & $39.63_{-2.97}^{+4.66}$ & $39.31_{-2.68}^{+4.72}$ & $39.06_{-2.63}^{+4.46}$ \\
				$m_2~{\rm [M_\odot]}$ & $32.53_{-3.77}^{+3.37}$ & $33.34_{-4.04}^{+3.06}$ & $33.66_{-3.82}^{+2.86}$ & $33.25_{-3.46}^{+3.10}$ & $33.83_{-3.88}^{+3.06}$ & $33.68_{-3.68}^{+2.90}$ \\
				$q$ & $1.22_{-0.19}^{+0.29}$ & $1.20_{-0.17}^{+0.27}$ & $1.15_{-0.13}^{+0.26}$ & $1.19_{-0.16}^{+0.26}$ & $1.16_{-0.14}^{+0.28}$ & $1.16_{-0.14}^{+0.27}$  \\
				$\chi_1$ & $0.01_{-0.19}^{+0.28}$ & $0.05_{-0.18}^{+0.30}$ & $0.01_{-0.18}^{+0.26}$ & $0.02_{-0.22}^{+0.37}$ & $0.02_{-0.17}^{+0.27}$ & $0.01_{-0.22}^{+0.23}$  \\
				$\chi_2$ & $-0.01_{-0.30}^{+0.24}$ & $0.02_{-0.25}^{+0.30}$ & $0.00_{-0.27}^{+0.29}$ & $0.03_{-0.29}^{+0.33}$ & $0.03_{-0.25}^{+0.31}$ & $0.02_{-0.21}^{+0.28}$  \\
				$\chi_{\rm eff}$ & $0.00_{-0.08}^{+0.10}$ & $0.06_{-0.09}^{+0.09}$ & $0.02_{-0.09}^{+0.11}$ & $0.05_{-0.10}^{+0.11}$ & $0.04_{-0.10}^{+0.11}$ & $0.03_{-0.10}^{+0.10}$  \\
				$D_{\rm L}~{\rm [Mpc]}$ & $471_{-185}^{+130}$ & $464_{-214}^{+143}$ & $495_{-179}^{+110}$ & $505_{-179}^{+112}$ & $549_{-161}^{+112}$ & $506_{-133}^{+124}$  \\
				$\iota~{\rm [rad]}$ & $2.62_{-0.56}^{+0.37}$ & $2.53_{-0.62}^{+0.42}$ & $2.60_{-0.58}^{+0.37}$ & $2.68_{-0.40}^{+0.32}$ & $2.74_{-0.50}^{+0.29}$ & $2.70_{-0.40}^{+0.31}$  \\
				$\alpha~{\rm [rad]}$ & $1.88_{-0.84}^{+0.70}$ & $1.99_{-1.00}^{+0.60}$ & $2.13_{-0.85}^{+0.46}$ & $2.01_{-1.02}^{+0.56}$ & $2.12_{-1.01}^{+0.45}$ & $1.82_{-0.71}^{+0.74}$ \\
				$\delta~{\rm [rad]}$ & $-1.23_{-0.05}^{+0.24}$ & $-1.22_{-0.06}^{+0.25}$ & $-1.23_{-0.05}^{+0.21}$ & $-1.23_{-0.06}^{+0.22}$ & $-1.22_{-0.06}^{+0.19}$ & $-1.24_{-0.04}^{+0.20}$ \\
				\hline
				${\rm log}\mathcal{B}^{\rm S}_{\rm N}$ & $286.10 \pm 0.15$ & $285.27 \pm 0.15$ & $285.15 \pm 0.15$ & $285.12 \pm 0.16$ & $285.44 \pm 0.16$ & $285.10 \pm 0.16$ \\	
			\end{tabular}
		\end{ruledtabular}
	\end{center}
\end{table*}

\subsection{Examples: EOBNR phasing and fluxes with and without NQC corrections}
\label{sbsec:nqcomp}
Before producing EOB/NR comparisons over the full database of NR simulations
used in Paper~I, let us discuss the effect of the various NQC choices on an illustrative
example. We choose configuration $(2,+0.85,+0.85)$, corresponding to SXS:BBH:0257-
\TEOB{} waveforms corresponding to this binary are generated with
three distinct options for NQC:
(i) the iterative procedure of Paper~I (here used with 4 iterations);
(ii) the new fits of Sec.~\ref{sbsec:newfits};
(iii) the absence of NQC parameters in the flux.
Figure~\ref{fig:nqc_test} illustrates the EOB/NR phase difference 
$\Delta\phi_{\rm 22}^{\rm EOBNR}\equiv \phi_{22}^{\rm EOB}-\phi_{22}^{\rm NR}$ 
for these three cases, plotted versus dimensionless time $t\equiv T/M$. 
The NQC parameters  are typically of order unity,
consistently with what pointed out in the test-mass limit (see in particular discussion around
Eq.~(12) of Ref.~\cite{Damour:2007xr}). For the iterated case, we have 
$(a_1^{(2,2)},a_2^{(2,2)})=(-0.2245,1.2917)$, while the fit consistently yields $(-0.2368,1.1964)$.
The EOB waveforms are aligned to the NR one by choosing relative time and phase shifts 
so to minimize the phase difference on  the dimensionless gravitational wave 
frequency interval $[M\omega_L,M\omega_R]=[0.034,0.045]$. The corresponding temporal
interval is indicated by the dash-dotted vertical lines in the left panel of the plot.
The fitted NQC parameters deliver a waveform that is perfectly consistent (though not strictly identical) 
with the one obtained via the iterative procedure. 
For each one of the three cases, the maximum EOB/NR unfaithfulness $\max({\bar{F}})$ 
computed in the next section using Eq.~\eqref{eq:barF} is $0.414\%$,  $0.456\%$ and $1.7\%$.
Note that this last number corresponds to an accumulated phase difference $\sim 4$~rad around 
merger time.

The presence of iterated NQC correction is also essential to yield consistency between
the NR angular momentum flux and the EOB flux, i.e. the radiation reaction force, with the opposite sign, 
that drives the inspiral dynamics. Figure~\ref{fig:flux_comparison} demonstrates this
fact for a specific dataset. A more detailed and systematic analysis will
will be discussed elsewhere~\cite{Albertini:2021}. To our knowledge, this is the 
first EOB/NR flux comparison after earlier work~\cite{Boyle:2008ge}. This analysis
is essential to cross check the reliability of radiation reaction, an approach that is 
well consolidated in the test-particle limit~\cite{Damour:2007xr,Bernuzzi:2010xj,Bernuzzi:2011aj,Harms:2014dqa}.
For comparable masses, it has never been exploited systematically because of 
the difficulty of computing it accurately from NR simulations~\cite{Boyle:2008ge}. 
Figure~\ref{fig:flux_comparison} demonstrates that, at least for the most recent SXS datasets, 
this is actually possible. The top panel of the figure shows Newton-normalized
angular momentum fluxes, while the bottom panel the EOB/NR fractional differences.
Specifically, we use $\dot{J}_{\rm Newt}^{\rm circ}=32/5\nu^2 x^{7/2}$, where we define
the frequency parameter from the GW quadrupole frequency $\omega_{22}$ as
$x\equiv (\omega_{22}/2)^{2/3}$. Note that $\omega_{22}=\omega_{22}^{\rm EOB,NR}$
in the EOB or NR case. The figure reports: (i) the raw NR angular momentum flux
summed over all multipoles up to $\ell_{\rm max}=8$; (ii) the smoothed one, where the high-frequency
noise (see inset) related to residual eccentricity and extrapolation has been eliminated with 
a specific fitting procedure~\cite{Albertini:2021}; (iii) the EOB flux, summed up to $\ell_{\rm max}=8$, 
{\it with} the iterated NQC correction factor in the flux, as described in Ref.~\cite{Damour:2009kr}; 
(iv) the same {\it without} the NQC correction factor. The top panel of Fig.~\ref{fig:flux_comparison} also
display the 3.5~PN accurate Taylor expanded flux along circular orbits. The vertical lines mark
the EOB Last Stable Orbit (LSO) as well as the location of the NR merger. It is important to note that
this comparison does not depend on an arbitrary time and phase shift (as it happens in waveform comparisons).
It is an intrinsic observable, complementary to the energy/angular momentum curves~\cite{Nagar:2015xqa,Damour:2011fu}, 
that in principle could be used to improve the current knowledge of the resummed analytical flux.
When looking at fractional differences (bottom panel) one sees that the inclusion of NR-informed 
NQC corrections in the flux yields a EOB/NR agreement at the level of the NR uncertainty up to the LSO location. 
The uncertainty on the NR data  is obtained, as usual, by taking the fractional difference between 
the highest and second highest resolutions available. Incorporating  NR-informed NQC corrections in 
the flux is thus an essential building element of \TEOBResumS{}, since it guarantees the physical 
correctness of the (self-consistent) EOB dynamics driven by radiation reaction.

\section{EOB/NR unfaithfulness} 
\label{sec:F}

Paper~I assessed the quality of the $(2,2)$ mode of \TEOB{} by comparing it to 
a total set of 595 SXS and 19 BAM waveforms. Each EOB waveform was generated
using 4 to 5 iterations. The overall comparison was done computing the EOB/NR unfaithfulness
 $\bar{F}(M)$ as a function of the total mass $M$. The unfaithfulness $\bar{F}$ between 
two waveforms $(h_1,h_2)$ is defined by
\be
\label{eq:barF}
\bar{F} \equiv  1- F=1- \max_{t_c, \phi_c} \frac{(h_1, h_2)}{\sqrt{(h_1,h_1)(h_2,h_2)}} \ ,
\ee
where $t_c$ and $\phi_c$ denote the time and phase at coalescence, 
and the Wiener scalar product
associated to the power-spectral density (PSD) of the
detector, $S_n(f)$, is
$(h_1, h_2) := 4\ \Re{ \int_{f_{\rm max}}^{f_{\rm max}}\,\mathrm{d}f\, \tilde{h}_1^(f)\tilde{h}_2^*(f)/S_n(f)}$,
where $\tilde{h}_1(f)$ is the Fourier transform of $h_1(t)$.
For the computation of EOB/NR unfaithfulness we use $f_{\rm min}$ as
the minimum NR frequency, and the Advanced LIGO PSD~\cite{Sn:advLIGO}.
The full EOB/NR unfaithfulness calculations of Paper~I was shown in Figs.~3 and 4 
therein (and it is shown again in Fig.~\ref{fig:barF_iter} for completeness): it 
is always below $0.5\%$ except for a single outlier that reaches the $0.85\%$.
Here we repeat such calculation, but with important differences: (i) we use the fits determined 
in the section above for $(a_1^{(2,2)},a_2^{(2,2)})$, so that we do not have to iterate on the dynamics 
but still we have an improved consistency between the waveform and the flux; (ii) we  use the 
post-adiabatic approximation~\cite{Nagar:2018gnk} to efficiently compute the inspiral part. 
The PA dynamics is computed at the 8th PA order on a grid with separation $dr = 0.1$ and 
stops at $r = 14$. The other EOB parameters are the same as in Paper~I and corresponds to the 
default configuration of \TEOB{} for parameter estimation.
In addition we also compute $\bar{F}$ {\it without} the NQC correction in the flux. The results are 
summarized in Fig.~\ref{fig:barF} without fits in the top row and {\it with fits} in the bottom row.
Each figure collects four panels that refer to different subsets of the NR simulations available,
separated according to the convenient classification of Paper~I. From left to right, each column
of the figure uses: spin-aligned SXS waveforms publicly released before February 3, 2019; spin-aligned 
SXS waveform data publicly released after February 3, 2019; spin-aligned BAM data; nonspinning SXS
and BAM data, up to mass ratio $q=18$.
The absence of the NQC corrections in radiation reaction increases $\max(\bar{F}$) up to
(a still acceptable) $\sim 3\%$; by contrast, when the NQC fits are included one easily gets 
$\max(\bar{F})$ well below $1\%$, consistently with the results of the iteration. The global picture 
is summarized in Fig.~\ref{fig:histo} that highlights in a  single figure the improvement brought by the fits.

\section{Computational efficiency}
\label{sec:timing}

\begin{figure}[t]
	\begin{center}
		\includegraphics[width=0.45\textwidth]{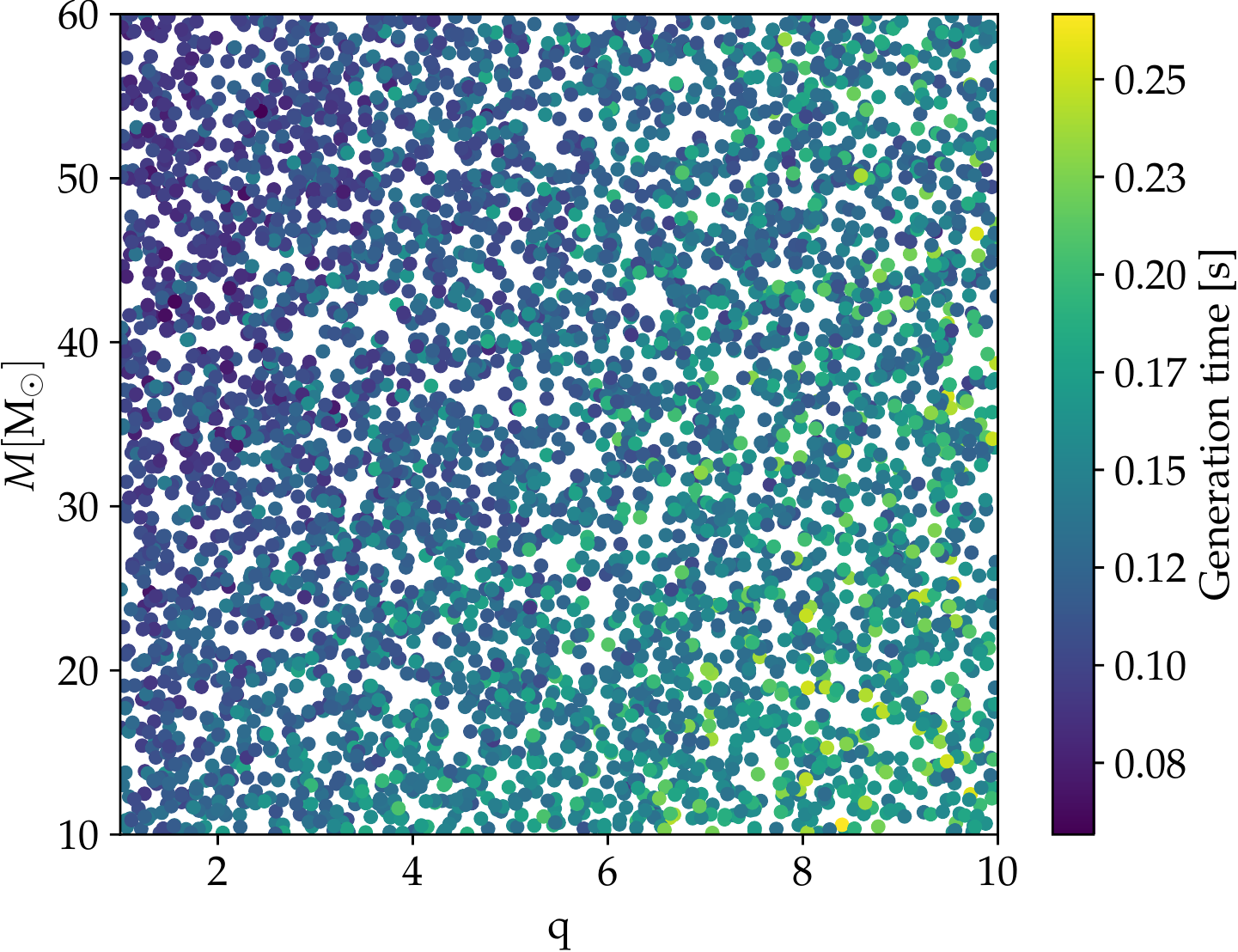}
		\includegraphics[width=0.45\textwidth]{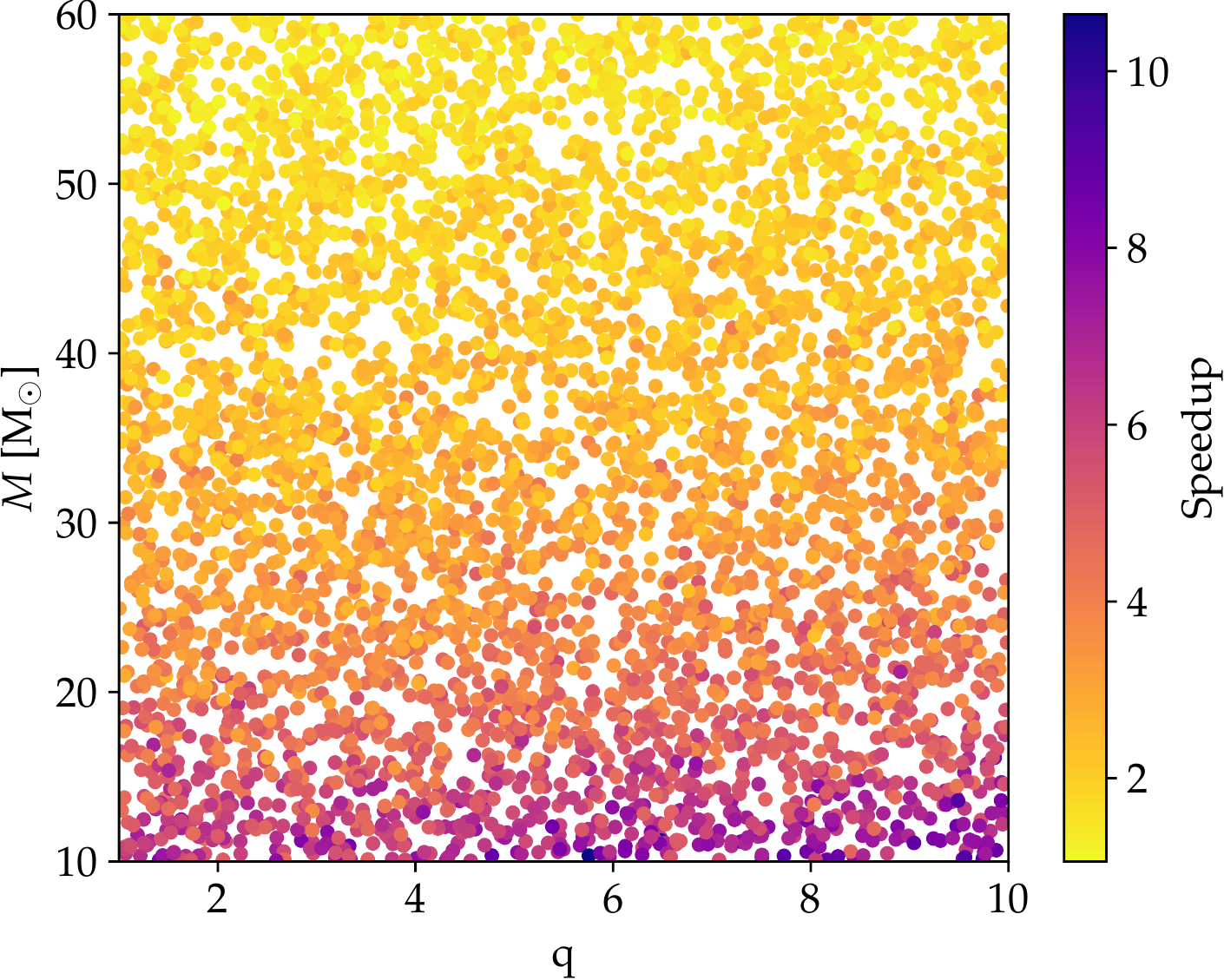}
		\caption{
		\label{fig:PAtime}
		\TEOB{} generation time for 5000 time-domain waveforms \textit{without} final interpolation on a uniform grid.
		The configurations are randomly sampled in $1 < q < 10$, $10 < M [M_\odot] < 60$ and $-1 < \chi_i < 1$ with starting GW frequency $f_0 = 10~{\rm Hz}$.
		Top panel: Computation time using the PA approximation to compute the dynamics.
		Bottom panel: Speed-up with respect to the same systems solving the complete ODEs.
		}
	\end{center}
\end{figure}

\begin{figure}[t]
	\begin{center}
		\includegraphics[width=0.47\textwidth]{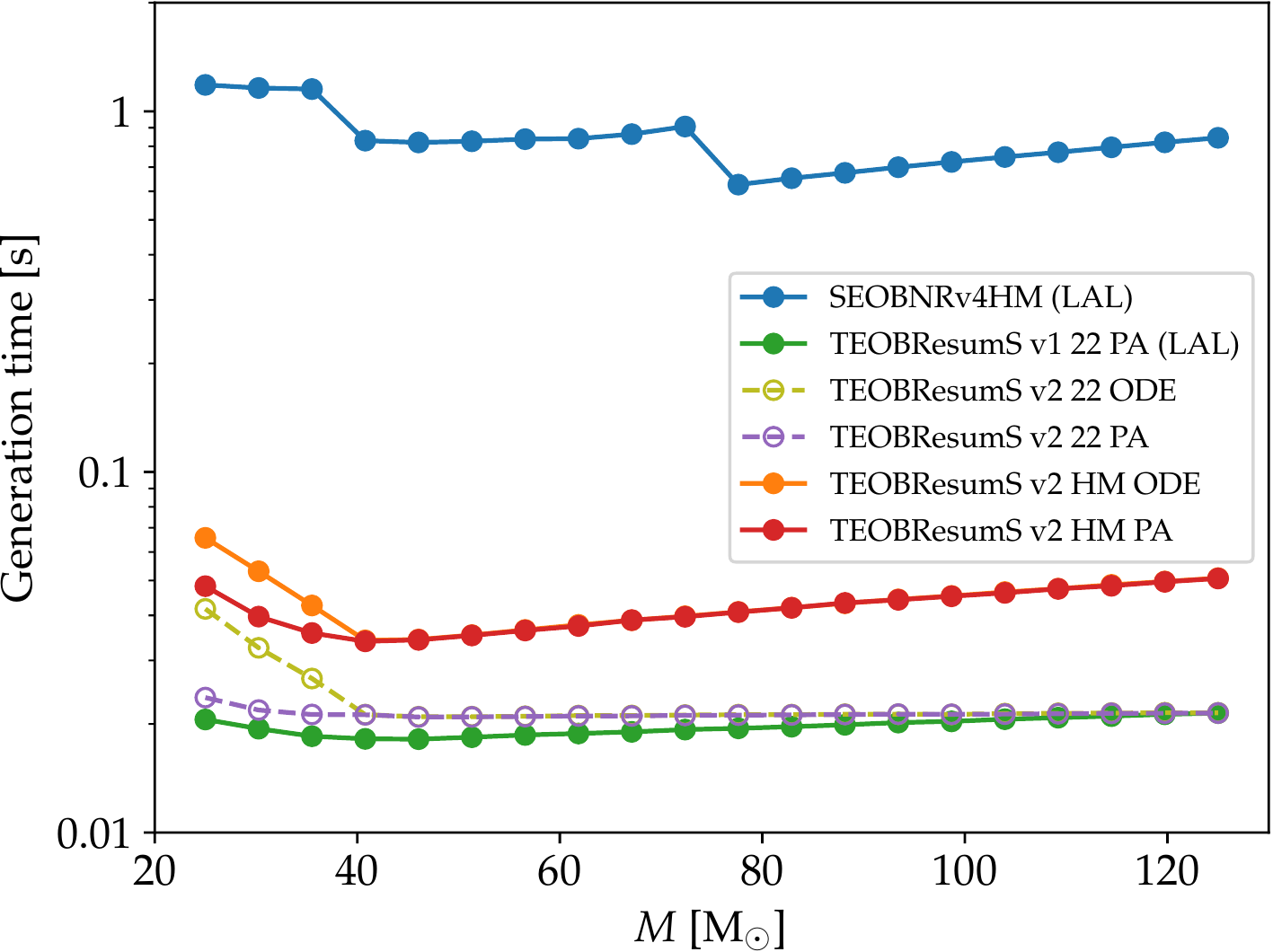}
		\includegraphics[width=0.47\textwidth]{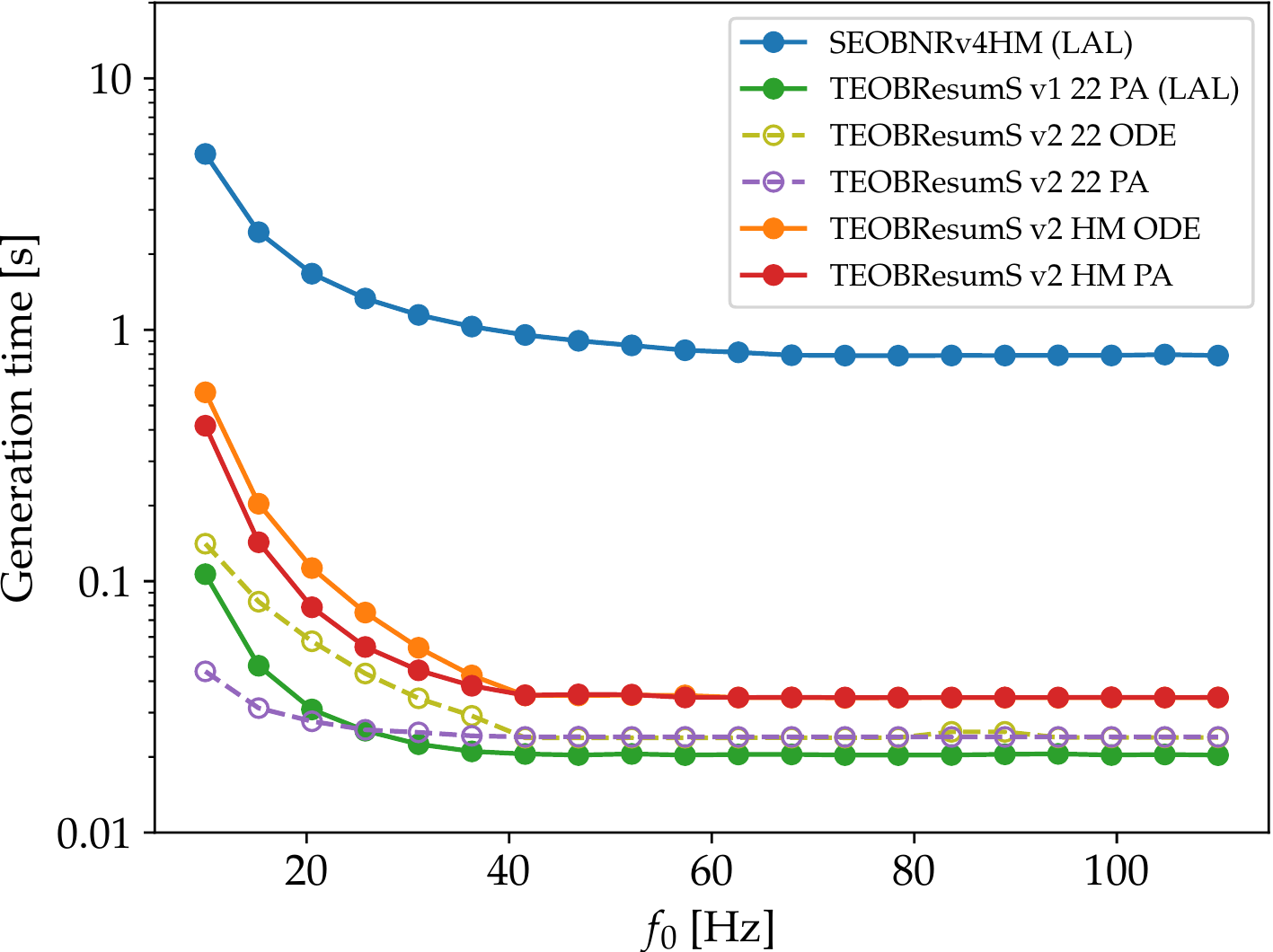}
		\caption{\label{fig:timing}
			\TEOB{} generation time of time-domain waveforms interpolated on a uniform-in-time grid with rate equal to 4096 Hz.
			We show the most recent iteration, \TEOB{} {\tt v2}, using either only the $\ell=m=2$ mode (22) or all the modes up to $\ell=8$ (HM).
			This is also run both employing the PA approximation (PA) or solving the full ODEs (ODE).
			We also show, as a comparison, the results for the LAL implementations of {\tt SEOBNRv4HM} and \TEOB{}.
			The latter, \TEOB{} {\tt v1}, uses the PA and does not include higher modes.
			The shown configurations correspond to $(q,\chi_1,\chi_2)=(2,+0.50,-0.30)$.			
			Top panel: 20 waveforms with starting frequency $f_0 = 30~{\rm Hz}$ and total mass $25M_\odot\leq M\leq 125{\rm M}_\odot$.
			Bottom panel: 20 waveforms with $M = 30~{\rm M_\odot}$ and varying $f_0$ between 10 and 110 Hz.
		}
	\end{center}
\end{figure}

In this Section we show the performance of \TEOB{} using the PA approximation \cite{Nagar:2018gnk}. 
The latter is used to avoid part of the computation of Hamilton's equations, that in the case of a nonprecessing 
system consist of 4 ordinary differential equations (ODEs).
Its use can be extended to any EOB-based model, as shown in Sec. VI of Ref.~\cite{Rettegno:2019tzh}. 
Within \TEOB{}, the 8th PA order is generally used to compute the radial and angular momenta on a radial grid, 
starting at the initial radius $r_0$, ending at dimensionless separation $r = 14$, with step $dr = 0.1$.
The other two dynamical variables, time and phase, are then calculated through an integration on the radial grid, 
essentially halving the number of necessary integrations. Beyond $r = 14$ the approximation could become unreliable for certain
configurations and hence the full ODEs are solved in the usual way\footnote{For simplicity, we ended the PA at $r=14$ as a robust,
conservative, choice all over the parameter space. This limit could actually be fine tuned as a function of the binary spin content and
lowered below $r=10$~\cite{Nagar:2018gnk}.}
The computational gain of using the PA approximation to compute full
waveforms is preliminarily discussed in Appendices
of Refs.~\cite{Nagar:2018plt,Akcay:2018yyh}, we present here a more detailed set of results.

In Fig.~\ref{fig:PAtime}, we show the \TEOB{} waveform generation time and the speedup with respect to configurations when the 4 ODEs are solved for the whole evolution. 
As expected, the use of the PA approximation has a greater impact on longer waveforms (lower total mass).
We can also note that, even without this speedup, \TEOB{} is already fast in the context of EOB-based models.

To put these times into perspective, in Fig.~\ref{fig:timing} we compare \TEOB{} to its equivalent higher modes model of the {\tt SEOBNR} family, {\tt SEOBNRv4HM}~\cite{Bohe:2016gbl, Cotesta:2018fcv}.
This latter is implemented within the LIGO Algorithm Library (LAL) \cite{lalsuite,lal_approximants} and, the time of writing, does not employ the 
PA approximation.
The C implementation of \TEOB{}, denoted {\tt v2}, is run with different settings: using all the modes up to $\ell=8$ or just the 
dominant $\ell=m=2$ one; employing the PA approximation for the dynamics or solving the full ODEs.
These are compared to the LAL version of {\tt SEOBNRv4HM} and of the same \TEOB{}.
This older iteration, the \TEOB{} {\tt v1}, already employed the PA approximation, 
but did not include higher modes nor NQC corrections in the flux. As we can expect, models which 
only include the $\ell=m=2$ multipole are found to be faster.
At the same time, we can see that the PA approximation (that is never employed in \TEOB{} when systems would start at $r_0 < 14$) 
improves the performance for long waveforms.
When compared to {\tt SEOBNRv4HM}, we find that \TEOB{} is generally an order of magnitude 
faster\footnote{For a comparison of the two models differences in the conservative dynamics, and the application of the PA approximation to {\tt SEOBNRv4}, see Ref.~\cite{Rettegno:2019tzh}}.

We highlight that, in order to improve the {\tt SEOBNRv4HM} performances, a reduced 
order model in the frequency domain has been developed~\cite{Cotesta:2020qhw}, that accelerates the waveform generation time by a factor of 100-200.
In a similar effort, Ref.~\cite{Schmidt:2020yuu} has recently applied machine learning methods to 
both \TEOB{} and \SEOB{}~\cite{Bohe:2016gbl} and built time-domain models that achieve a speedup of 10 to 50 for \TEOB{} 
and about an order of magnitude more for \SEOB{}, see Fig.~7 of Ref.~\cite{Schmidt:2020yuu}.
This fact is consistent with our analysis of Fig.~\ref{fig:timing}: it reflects the difference 
in computational cost of the two baseline models.

In conclusion, our timing analysis indicates that the native implementation of \TEOB{} using
the PA approximation (including the {\tt v1} implementation distributed with LAL~\cite{lalsuite}) is efficient enough to be
used for parameter estimation, as we shall demonstrate in the following section.

\section{GW150914 analysis}
\label{sec:gw150914}

\begin{figure*}[t]
	\begin{center}
		\includegraphics[width=0.3\textwidth]{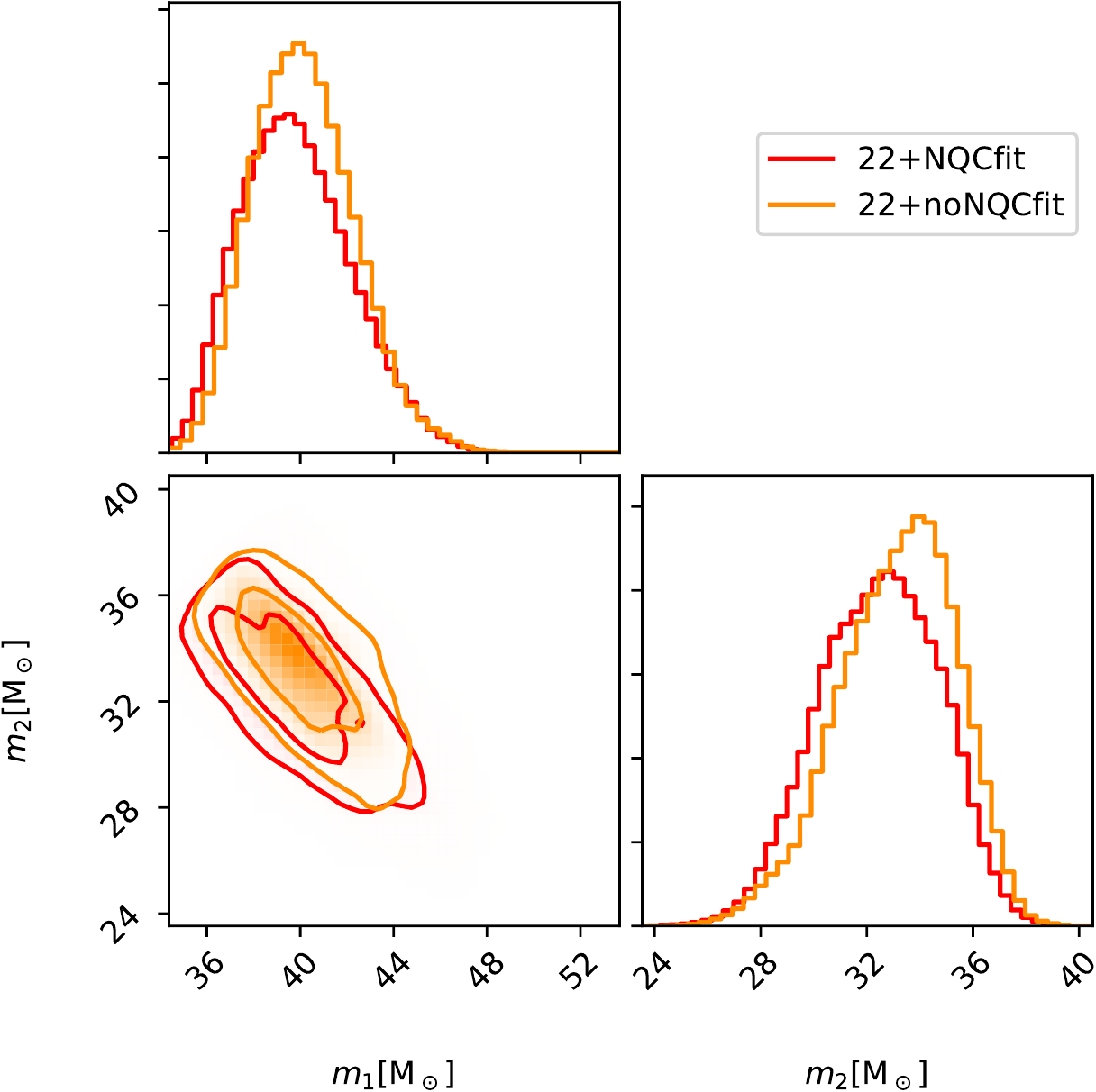}
		\includegraphics[width=0.3\textwidth]{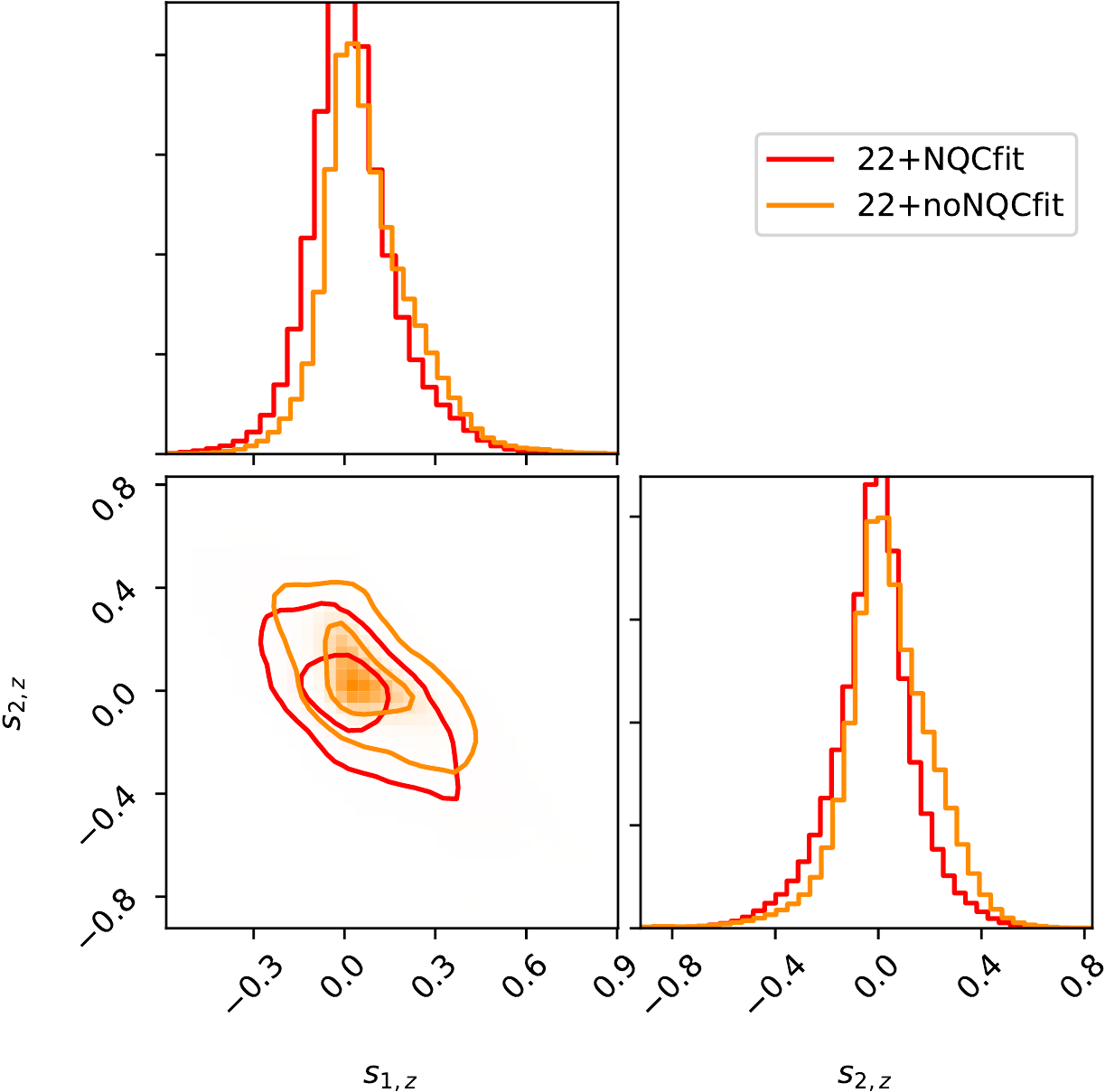}
		\includegraphics[width=0.3\textwidth]{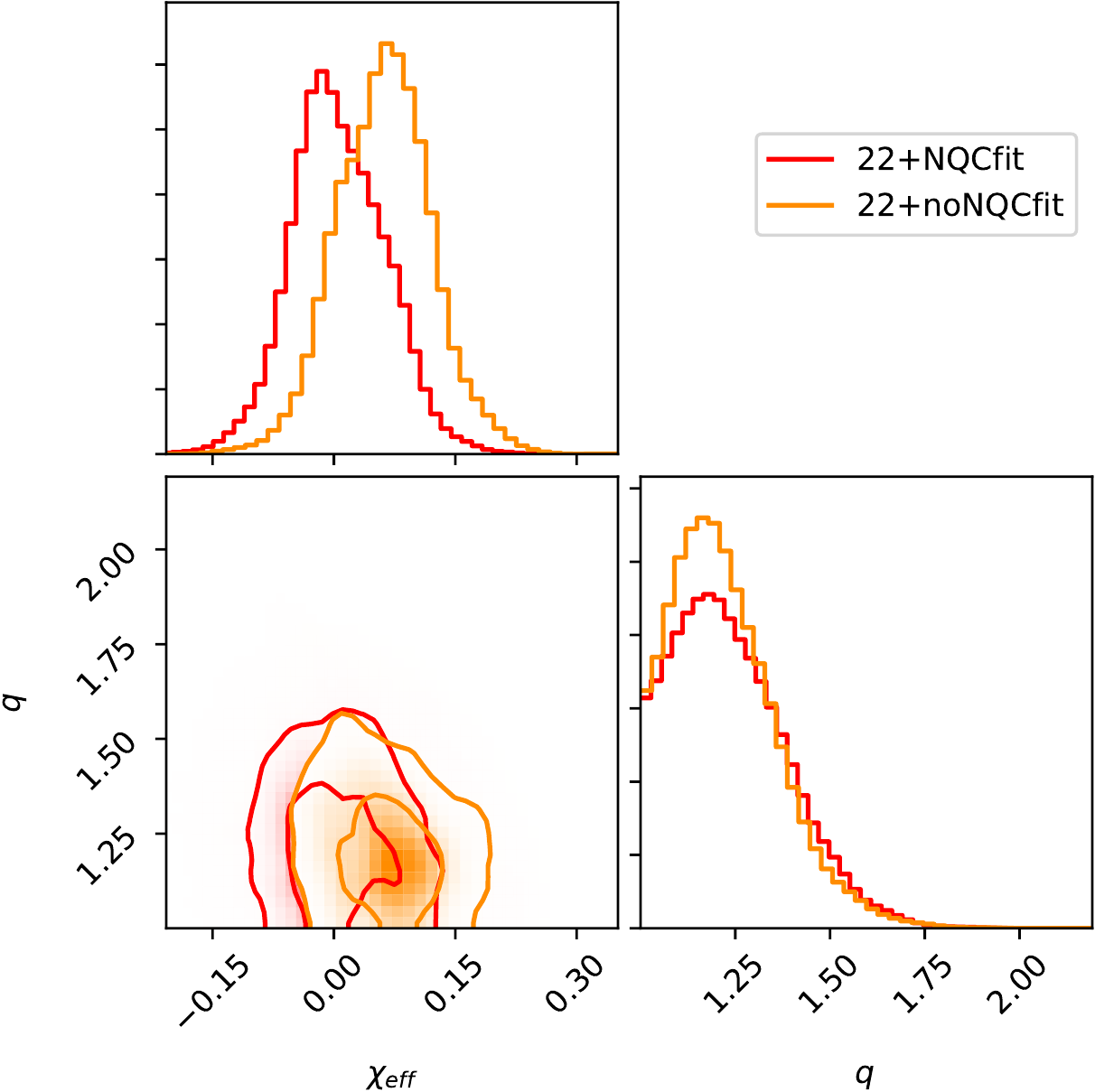}\\
		\includegraphics[width=0.3\textwidth]{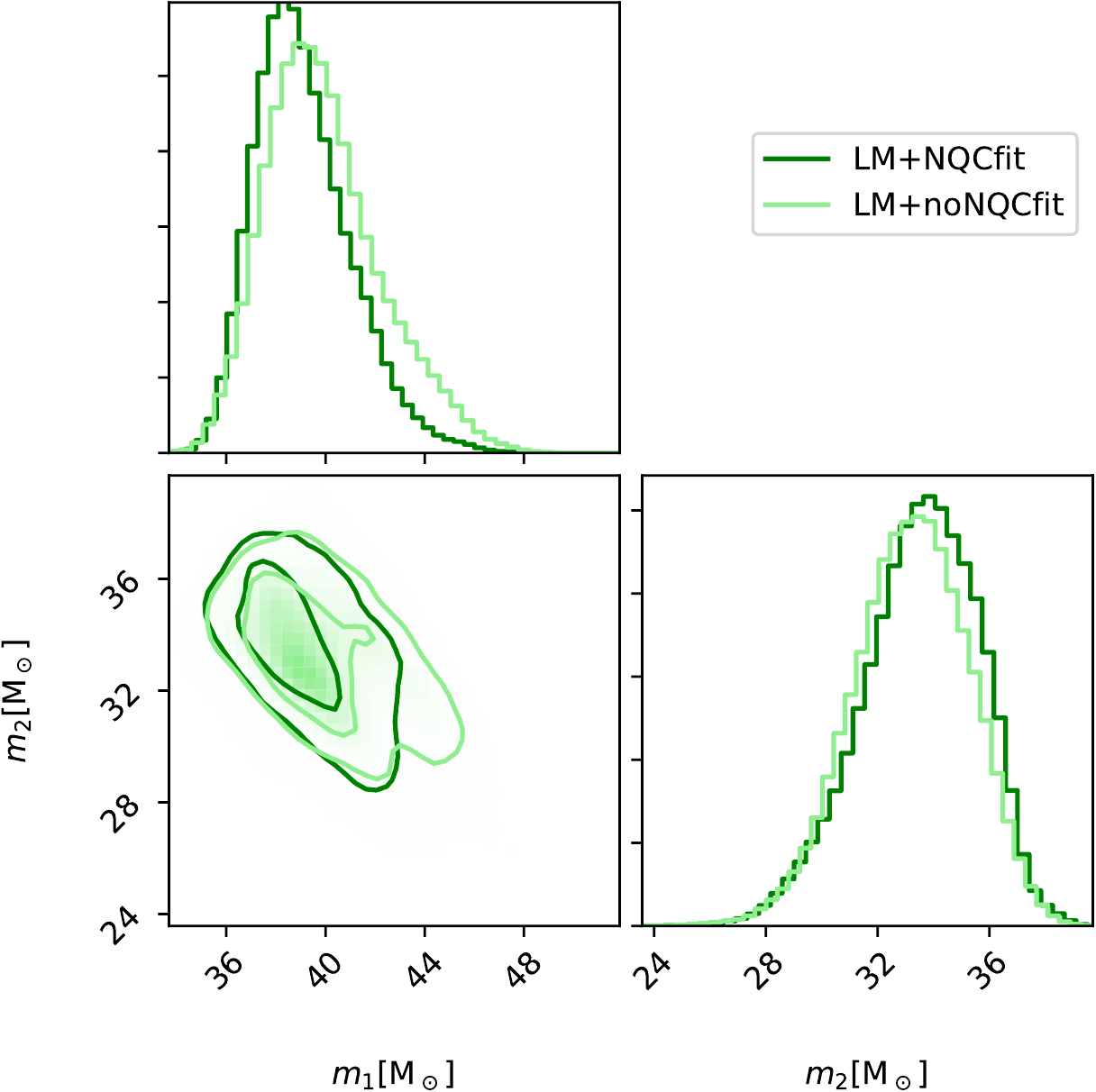}
		\includegraphics[width=0.3\textwidth]{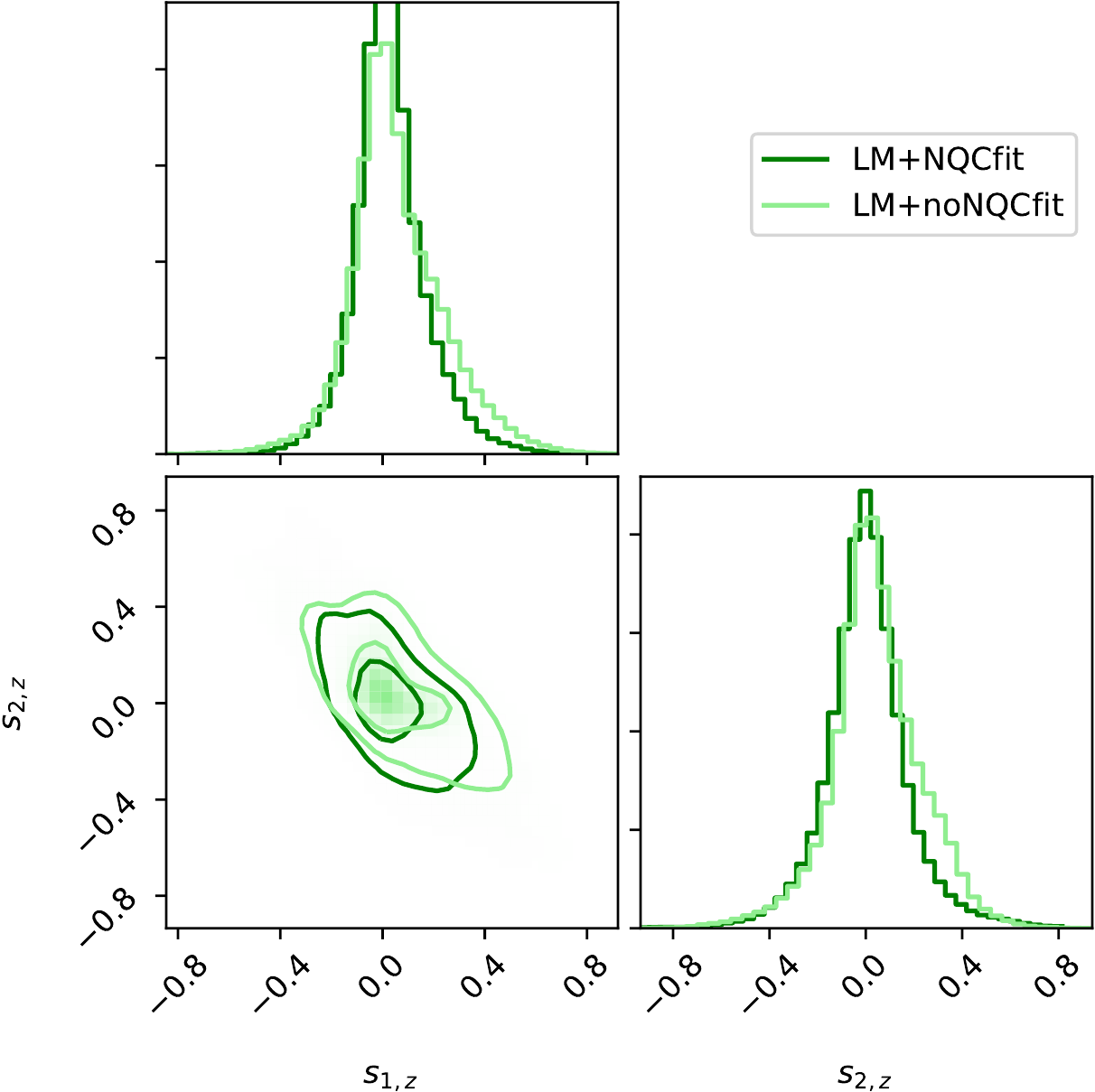}
		\includegraphics[width=0.3\textwidth]{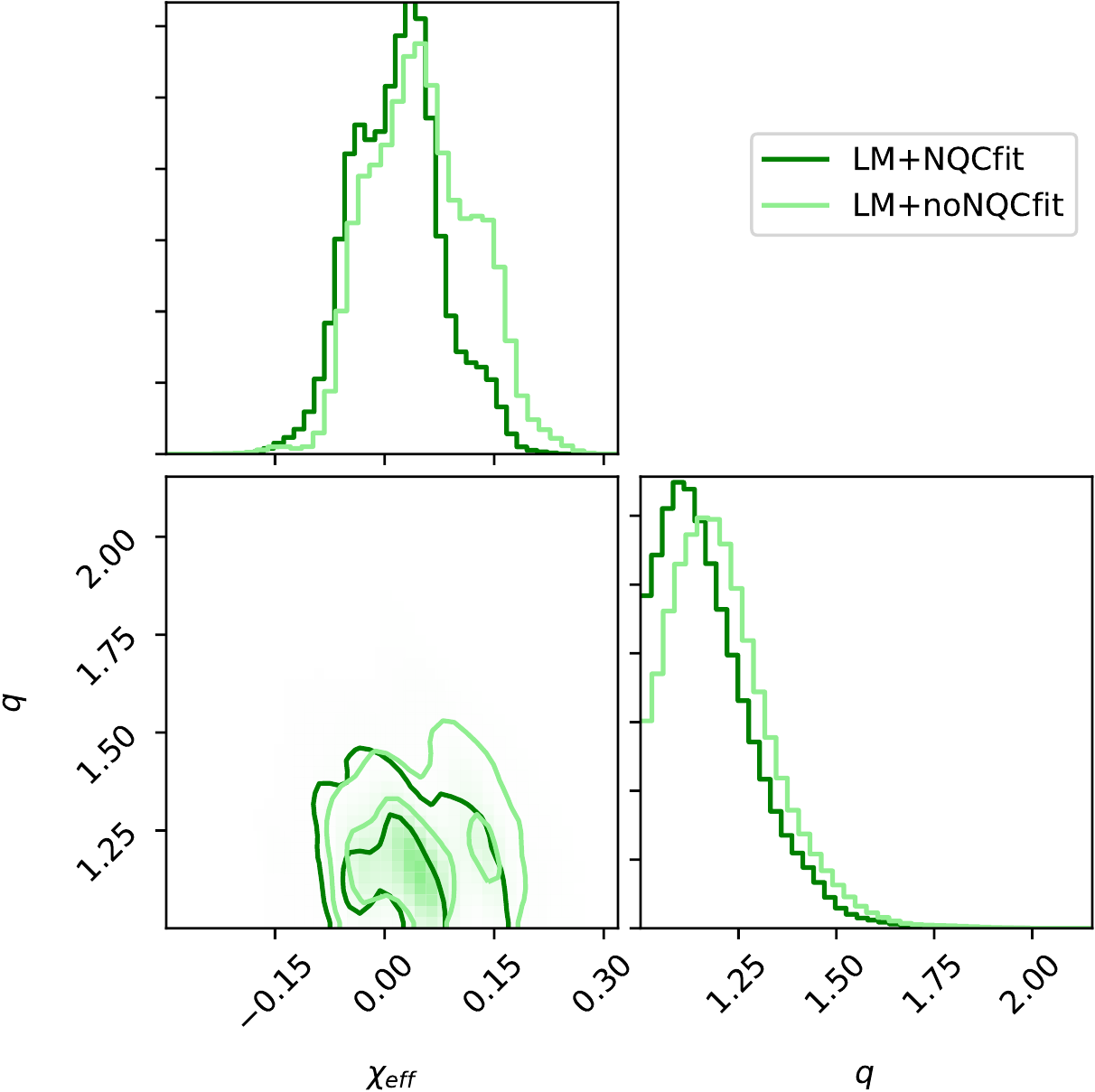}\\
		\includegraphics[width=0.3\textwidth]{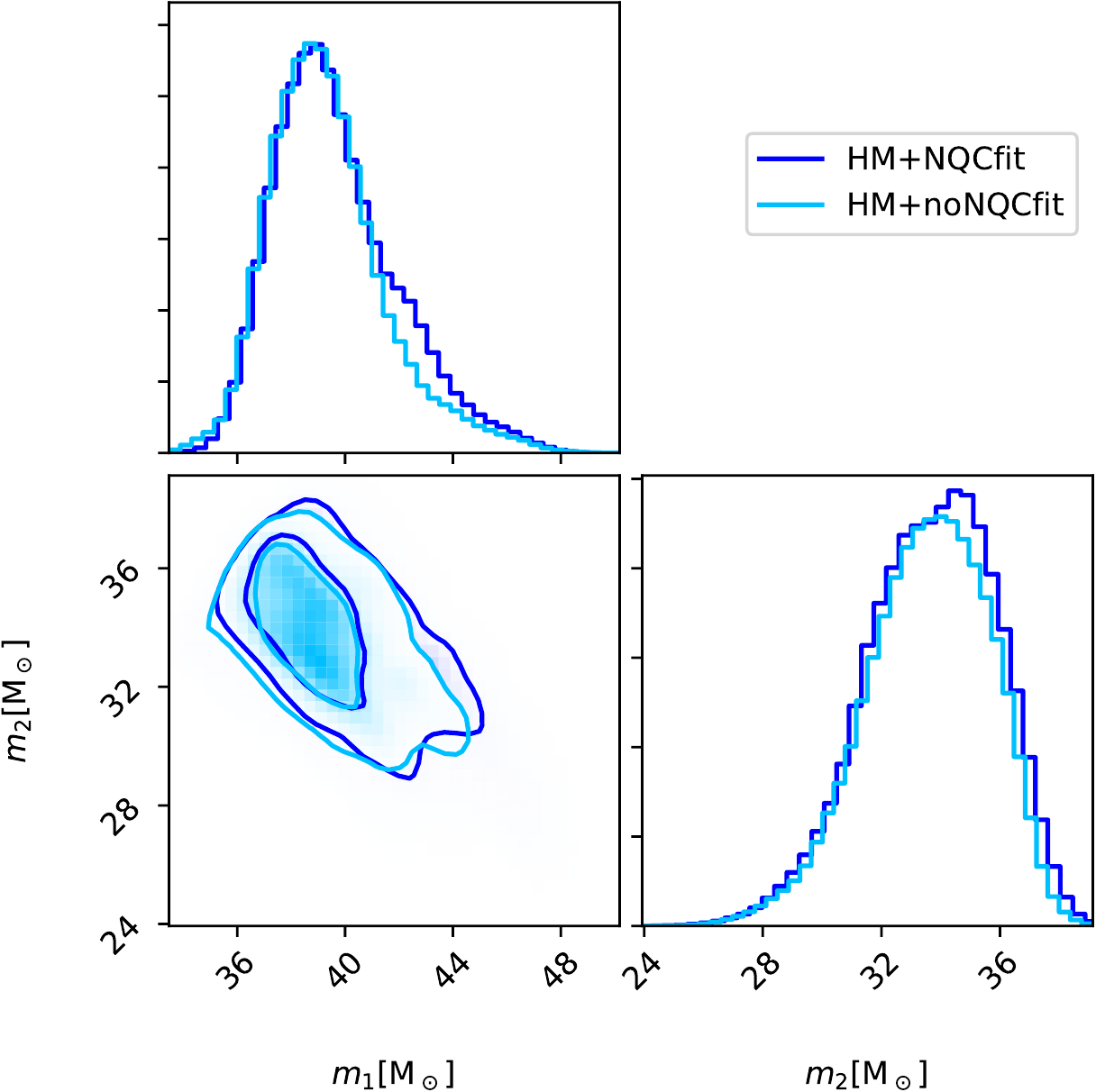}
		\includegraphics[width=0.3\textwidth]{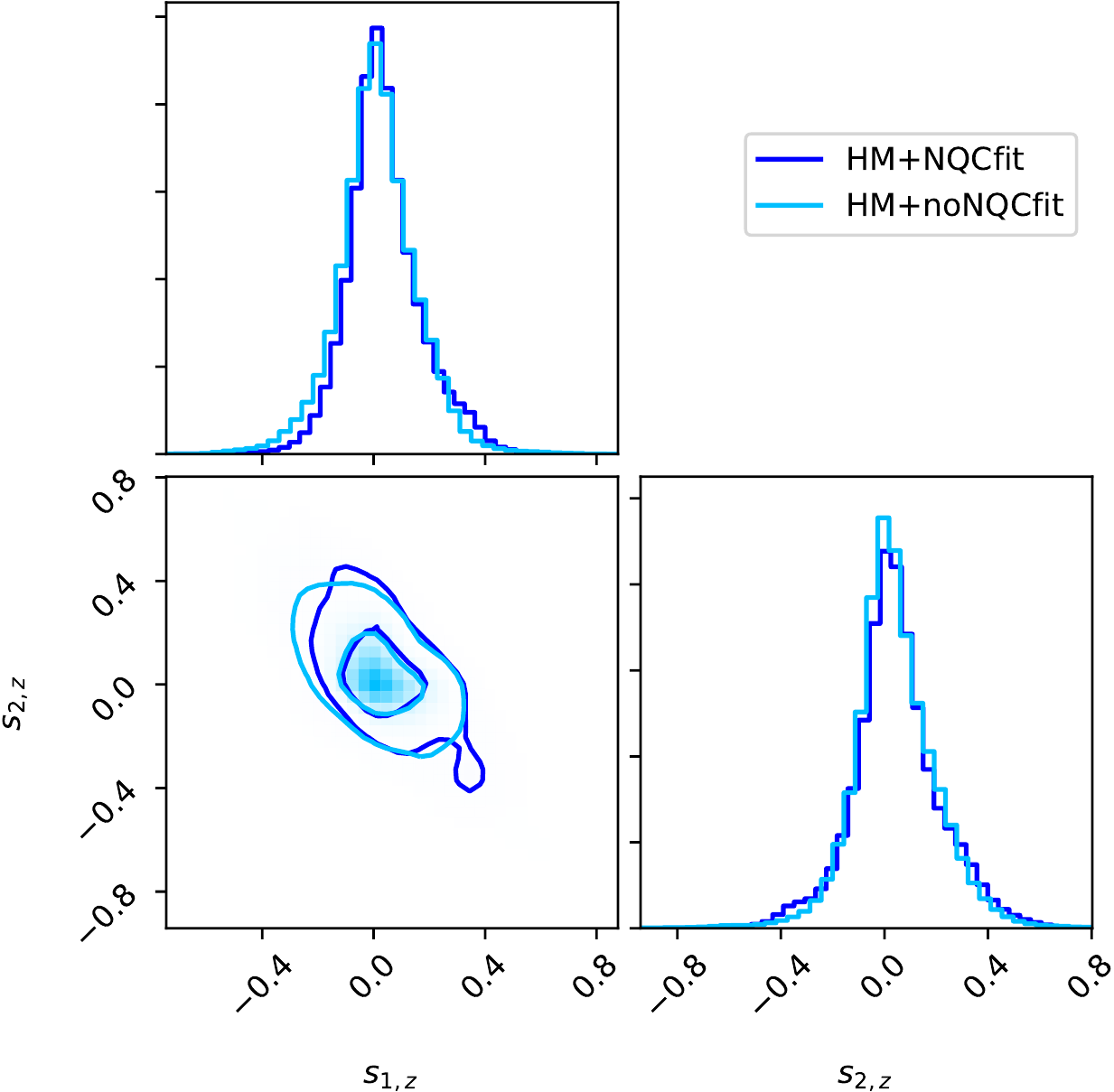}
		\includegraphics[width=0.3\textwidth]{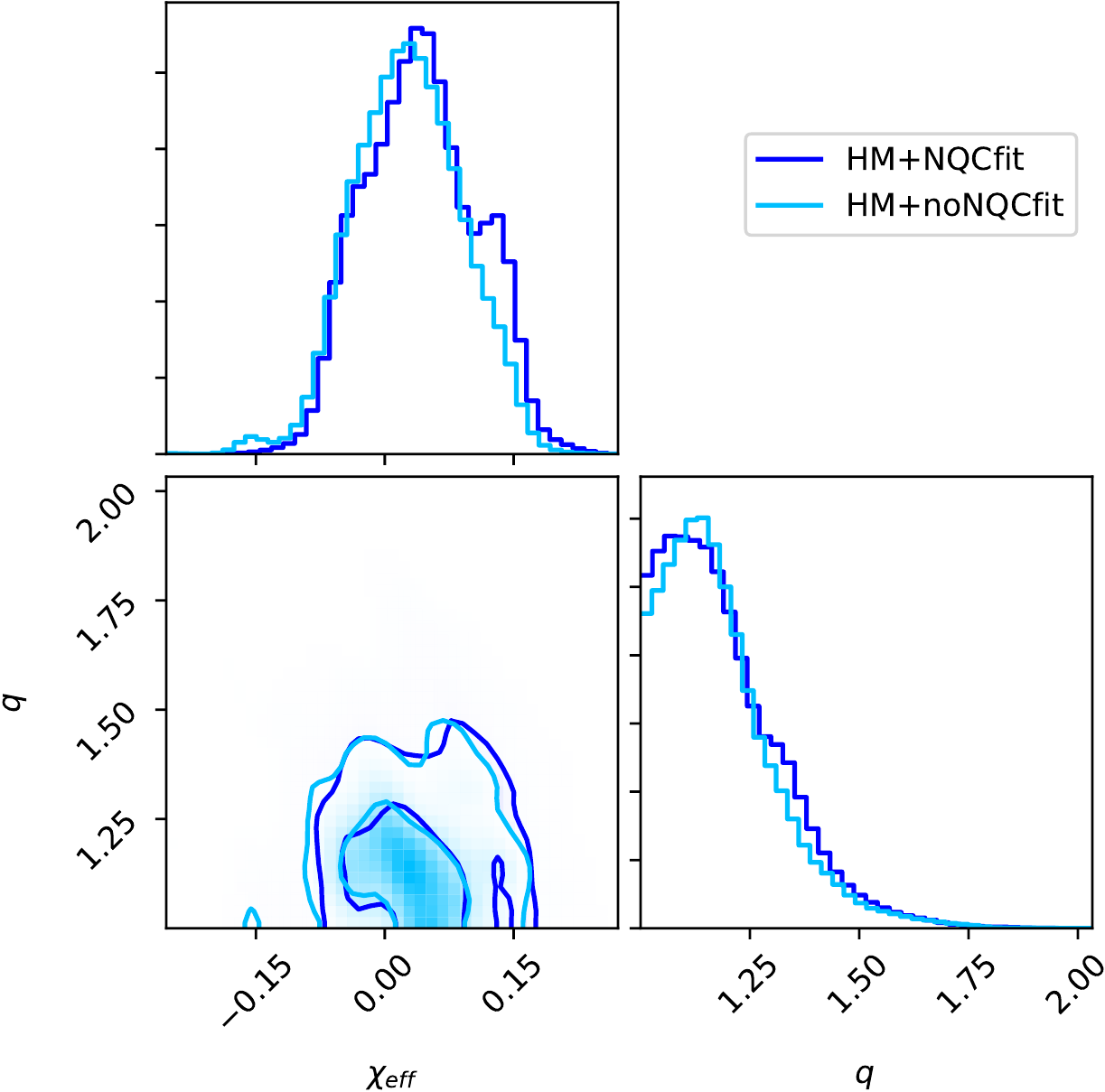}
		\caption{\label{fig:gw150914_fit_nofit}
			Parameter estimation of GW150914. Comparing parameter inference with and without NQC fits. 
			It is interesting to note that the effect of the NQC is highly subdominant when all the higher modes
			up to $\ell=8$ are included in the waveform.}
	\end{center}
\end{figure*}

\begin{figure}[t]
	\begin{center}
		\includegraphics[width=0.33\textwidth]{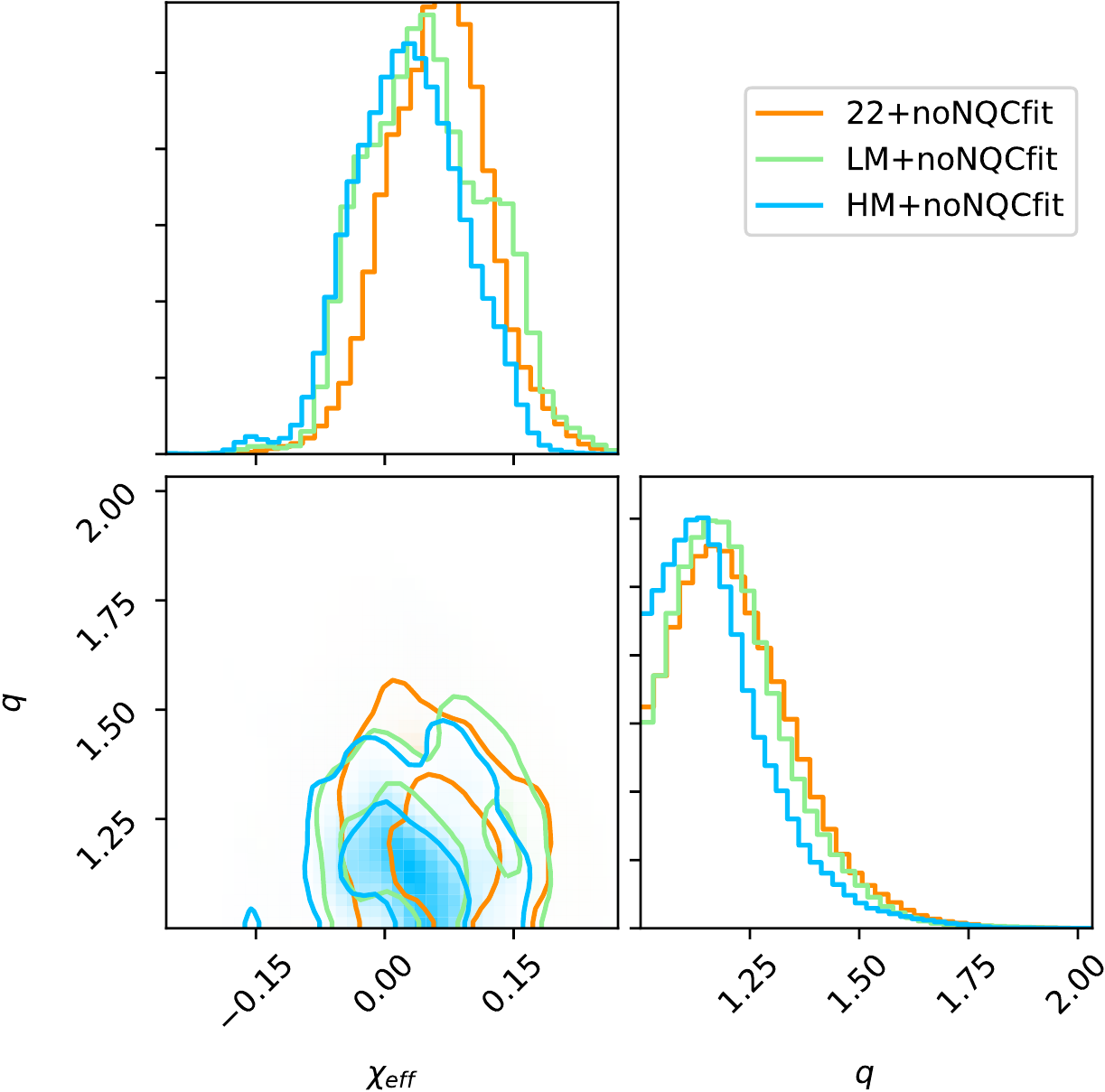}\\
		\includegraphics[width=0.33\textwidth]{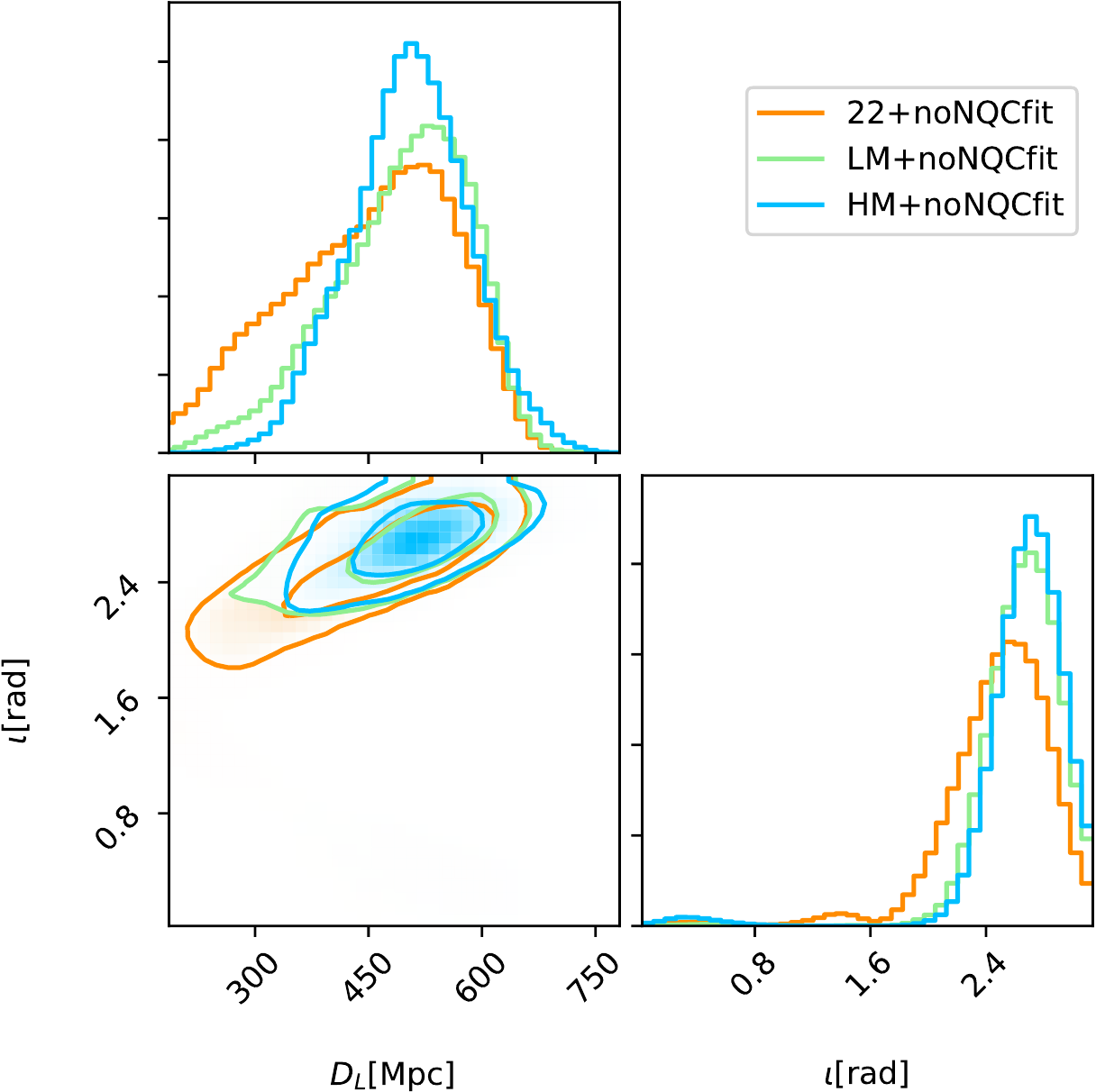}
		\caption{\label{fig:gw150914_nofit} 
			Parameter estimation of GW150914 without the NQC fits in radiation reaction.
			The figure compares posteriors of: (i) $\ell=m=2$ only waveform; (ii) multipolar waveform with all $\ell=m$ modes summed up to $\ell=5$; 
			(iii) complete waveform with all modes up to $\ell=8$.
		}
	\end{center}
\end{figure}

\begin{figure}[t]
	\begin{center}
		\includegraphics[width=0.33\textwidth]{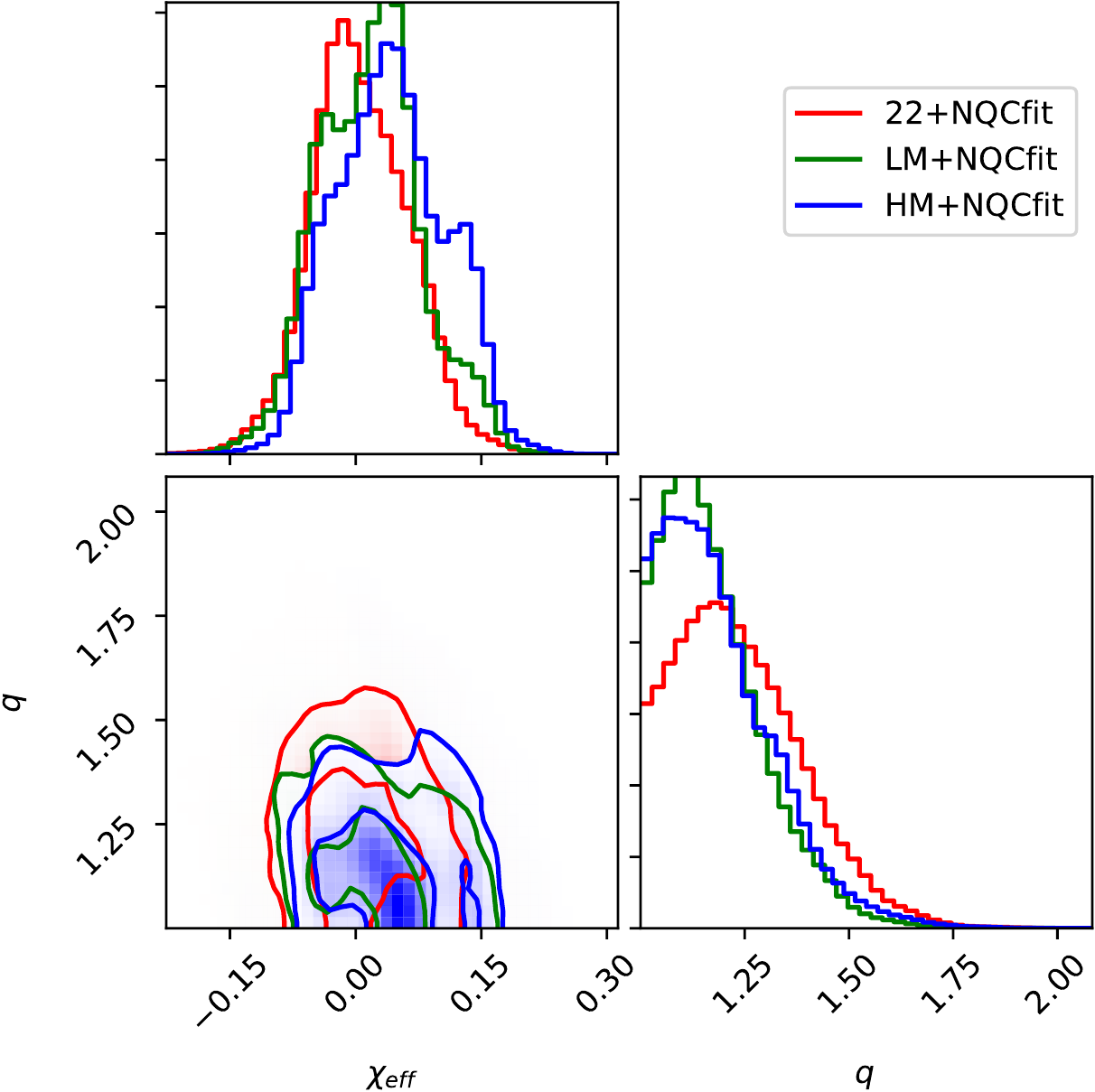}\\
		\includegraphics[width=0.33\textwidth]{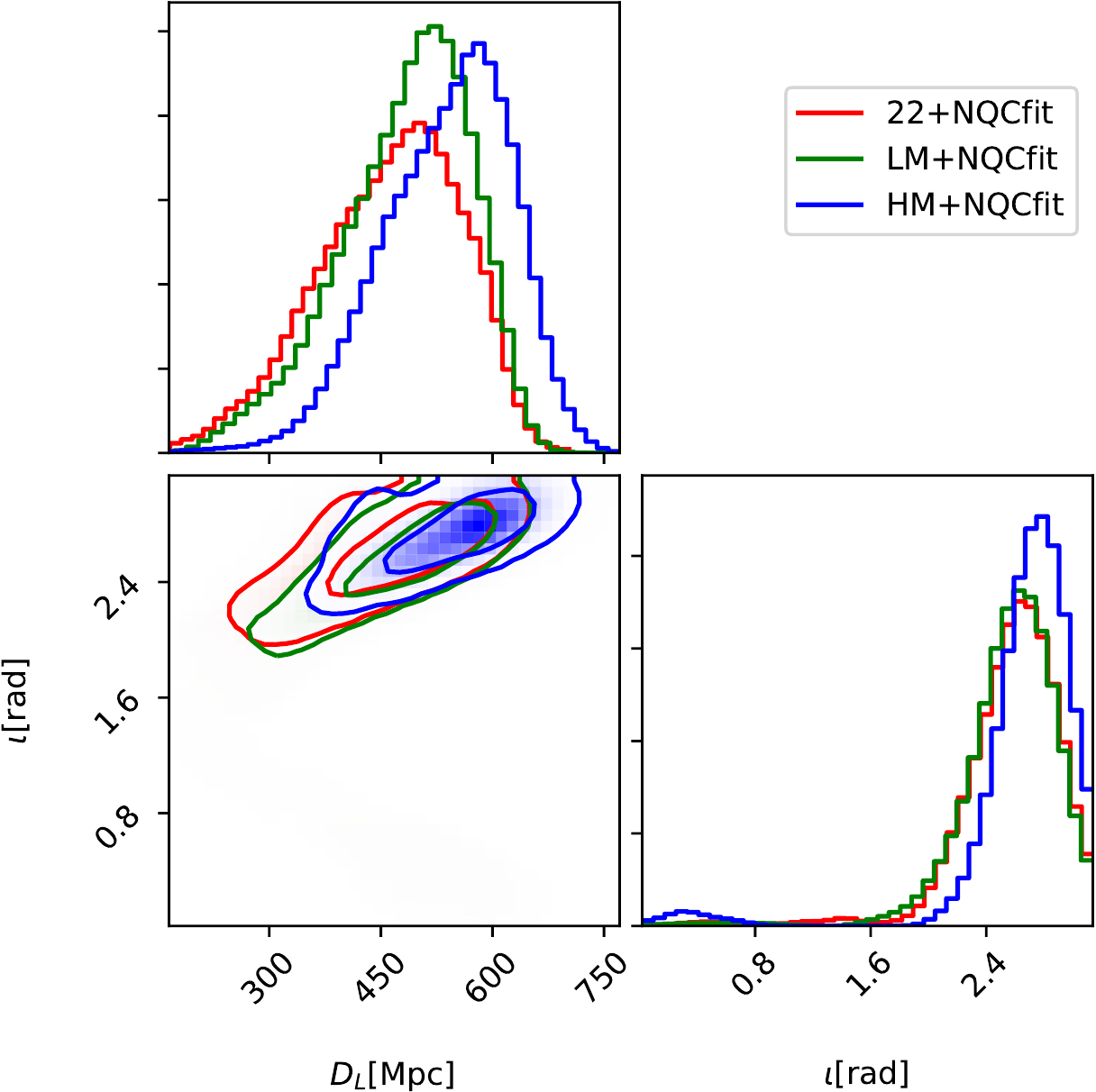}
		\caption{\label{fig:gw150914_fit}
			Parameter estimation of GW150914 with the NQC fits in radiation reaction.
			The figure compares posteriors of: (i) $\ell=m=2$ only waveform; (ii) multipolar waveform with all $\ell=m$ modes summed up to $\ell=5$; 
			(iii) complete waveform with all modes up to $\ell=8$.}
	\end{center}
\end{figure}

We ran a PE study on GW150914 using {\tt bajes}~\cite{Breschi:2021wzr}.
We employed the {\tt dynesty} sampler with 1024 live points and tolerance of 0.1.
We extracted the data from the GWOSC archive~\cite{Abbott:2019ebz} 
and analyzed $16$ seconds of data around GPS time $1126259462$, 
with a sampling rate of 4096 Hz in the range of frequencies $[20,1024]$~Hz.
We set the same prior distributions for all runs. 
The chirp mass prior was uniform in $[24,37] M_\odot$ and the mass ratio $q$ in $[1,8]$. 
We only considered aligned spins with an isotropic prior in the range $[-0.99,+0.99]$. 
We used a volumetric prior for the luminosity distance in $[100, 800]$~Mpc.

Separate runs are performed with \TEOBResumS{}, either including 
the new NQC fits in the radiation reaction or not. For each of the two cases, 
parameter estimation runs are performed with the $(2,2)$ mode only (22), 
the $\ell=m$ and $\ell \le 5$ modes (LM), and with all the modes up to
$\ell=m=8$ multipoles (HM). In this case, all the other subdominant 
modes except $(2,1)$, $(3,2)$, $(4,3)$ and $(4,2)$ do not use NR information
to be completed through merger and ringdown, but only rely on the analytical EOB
waveform (see e.g. Fig.~10 of Ref.~\cite{Nagar:2019wds}). 
We used the PA approximation of the dynamics for all runs, as it is the default 
option for our implementation (e.g. \cite{Gamba:2020wgg,Gamba:2020ljo,Breschi:2021wzr}.)
Each one of these analyses took about 2 days on 8 CPUs.
More details on the \TEOB computational cost can be found in Appendix \ref{sec:timing}.

The results of such runs are listed in Table~\ref{postresults}.
The difference of using the NQC fits is highlighted in Fig.~\ref{fig:gw150914_fit_nofit}.
Neglecting the NQC fits in the radiation reaction, that has a large impact on the EOB-NR unfaithfulness, 
has a very small effect on parameter estimation, despite the high SNR of GW150914. 
The only appreciable difference can be seen in the $\chi_{\rm eff}$ variable for the 22 run, 
which is more skewed towards 0 when NQC fits are used. 
It is interesting to note that the difference between using the NQC fits and not employing 
them tends to disappear when using more multipoles. 
Some effect in this direction was to be expected, since the NQC fits only affect the $\ell=m=2$ mode, 
which has a somewhat diminished importance when other multipoles are used.

Using the same data, we can attempt to determine whether this analysis is sensitive 
to the higher modes, given that the system is almost equal-mass and nonspinning.
There are no appreciable differences in the system parameters when using higher order multipoles, apart from a small preference for a mass ratio closer to 1.
Instead, using modes beyond the dominant $\ell=m=2$ one, helps to better constrain the source distance and inclination. 
In particular, the runs which employed a larger number of modes, seem to prefer larger distances and more face-on/away configuration.
These results are compatible with what found in Ref.~\cite{Kumar:2018hml} using the NR surrogates {\tt NRSur7dq2} and {\tt NRSur7dq2HM}.
This difference in posteriors is shown in Figs.~\ref{fig:gw150914_nofit} and \ref{fig:gw150914_fit}.

We conclude highlighting that using Bayes' factors, we cannot determine a preference 
for any of the models used for these analyses (see again Table~\ref{postresults}).

%

\section{Conclusion}
\label{sec:conc}

This work completes the description of the techniques employed in the
current \TEOB{} waveform (\texttt{v2})~\cite{Nagar:2018zoe,Nagar:2020pcj} and 
outlines a viable path towards the use of faithful EOB models in GW 
parameter estimation.
Here, we highlighted the importance of: (i) including NQC corrections in the
radiation reaction and (ii) using the post-adiabatic approximation to 
improve the computational efficiency of the inspiral.

The NQC fits developed here ensure an improved consistency between the
EOB dynamics (radiation reaction flux) and the waveform without the
need of an iterative procedure to determine the NQC parameters
$(a^{(2,2)}_1,a^{(2,2)}_2)$. The EOB/NR unfaithfulness achieved with 
this NQC setting and with the use of the post-adiabatic approximation to
the EOB dynamics is always below $0.01$, with 78.5\% of the 611 NR
waveforms below 0.001 (see right panel of Fig.~\ref{fig:histo}).

The PA approximation, together with an efficient implementation, makes
each version of \TEOB{} (including  \texttt{v1} distributed with LAL~\cite{lalsuite})
suitable for parameter estimation {\it in its native form}, without
the need of constructing surrogate or machine learning representations.
The latter can provide significant further speed
up~\cite{Schmidt:2020yuu}, but their construction becomes increasingly
more complex as more physics effects are included (spin precession,
eccentricity, etc). 

The application of \TEOB{} to GW150914, that still represents one 
of the highest signal-to-noise ratio event observed thus far, indicates 
that the present techniques are well suited for the unbiased analysis
of comparable-masses and moderately spinning binary black holes
signals. 
In particular, the analysis is not sensitive to the inclusion of NQC
fits in the radiation reaction, despite the  
inconsistency and far worse EOB/NR unfaithfulness of the model when these fits are not included.
The inclusion of higher modes beyond the $\ell=m=2$ one has an
appreciable effect only in giving a more stringent  
constraint of the source distance and inclination, as also seen with
NR surrogates~\cite{Kumar:2018hml}.

Future work should address the waveform systematics effects
and limitation of current EOB models for larger mass-ratio and/or
waveforms with larger spins. An important aspect in this respect, is
to explore phasing, faithfulness and full parameter estimation
altogether, as done for tidal effects in~\cite{Gamba:2020wgg}, in order
to identify which elements of the model require improvements and
the connection between the phasing and the parameter estimation.

The current techniques can be immediately applied to include
precessional effects~\cite{Akcay:2020qrj} and tides~\cite{Bernuzzi:2014owa,Akcay:2018yyh};
fast post-adiabatic multipolar waveforms with these features can be already generated with \TEOB{}.
The same computationally efficient infrastructure of \TEOB{} is also
shared by {\tt TEOBResumSGeneral}~\cite{Chiaramello:2020ehz,Nagar:2020xsk,Nagar:2021gss},  
that deals with either eccentric inspirals (although {\it without} the PA approximation) 
or hyperbolic scatterings.
Future work will also focus on rapid, and yet accurate, methods for
the solution of the eccentric EOB dynamics~\cite{Chiaramello:2020ehz,Nagar:2020xsk,Nagar:2021gss}, 
and on the extension of EOB to directly compute frequency-domain inspiral-merger-ringdown 
waveforms~\cite{Gamba:2020ljo}.

\acknowledgments
S.B. and M.B. acknowledge support by the EU H2020 under ERC Starting Grant, no.~BinGraSp-714626.  
Data analysis was performed on the supercomputer ARA at Jena. We
acknowledge the computational resources provided 
by Friedrich Schiller University Jena, supported in part by DFG grants
INST 275/334-1 FUGG and INST 275/363-1 FUGG.
Data postprocessing was performed on the Virgo ``Tullio'' server 
in Torino, supported by INFN.
{\tt TEOBResumS} is publicly available at
\url{https://bitbucket.org/eob_ihes/teobresums/}.
The {\tt v2} version of the code, that implements the PA approximation and higher modes, 
is fully documented in Refs.~\cite{Nagar:2018gnk,Nagar:2018plt,Nagar:2019wds,Nagar:2020pcj}
together with this work. We recommend the above references to be cited by \TEOBResumS{}
users.

\noindent
This research has made use of data, software and/or web tools obtained from the Gravitational Wave Open Science Center (https://www.gw-openscience.org/ ), a service of LIGO Laboratory, the LIGO Scientific Collaboration and the Virgo Collaboration. LIGO Laboratory and Advanced LIGO are funded by the United States National Science Foundation (NSF) as well as the Science and Technology Facilities Council (STFC) of the United Kingdom, the Max-Planck-Society (MPS), and the State of Niedersachsen/Germany for support of the construction of Advanced LIGO and construction and operation of the GEO600 detector. Additional support for Advanced LIGO was provided by the Australian Research Council. Virgo is funded, through the European Gravitational Observatory (EGO), by the French Centre National de Recherche Scientifique (CNRS), the Italian Istituto Nazionale di Fisica Nucleare (INFN) and the Dutch Nikhef, with contributions by institutions from Belgium, Germany, Greece, Hungary, Ireland, Japan, Monaco, Poland, Portugal, Spain.

\appendix

\section{NR faithfulnesses with NQC iterations}
\label{sec:F:iter}

\begin{figure*}[t]
  \begin{center}
    \includegraphics[width=0.32\textwidth]{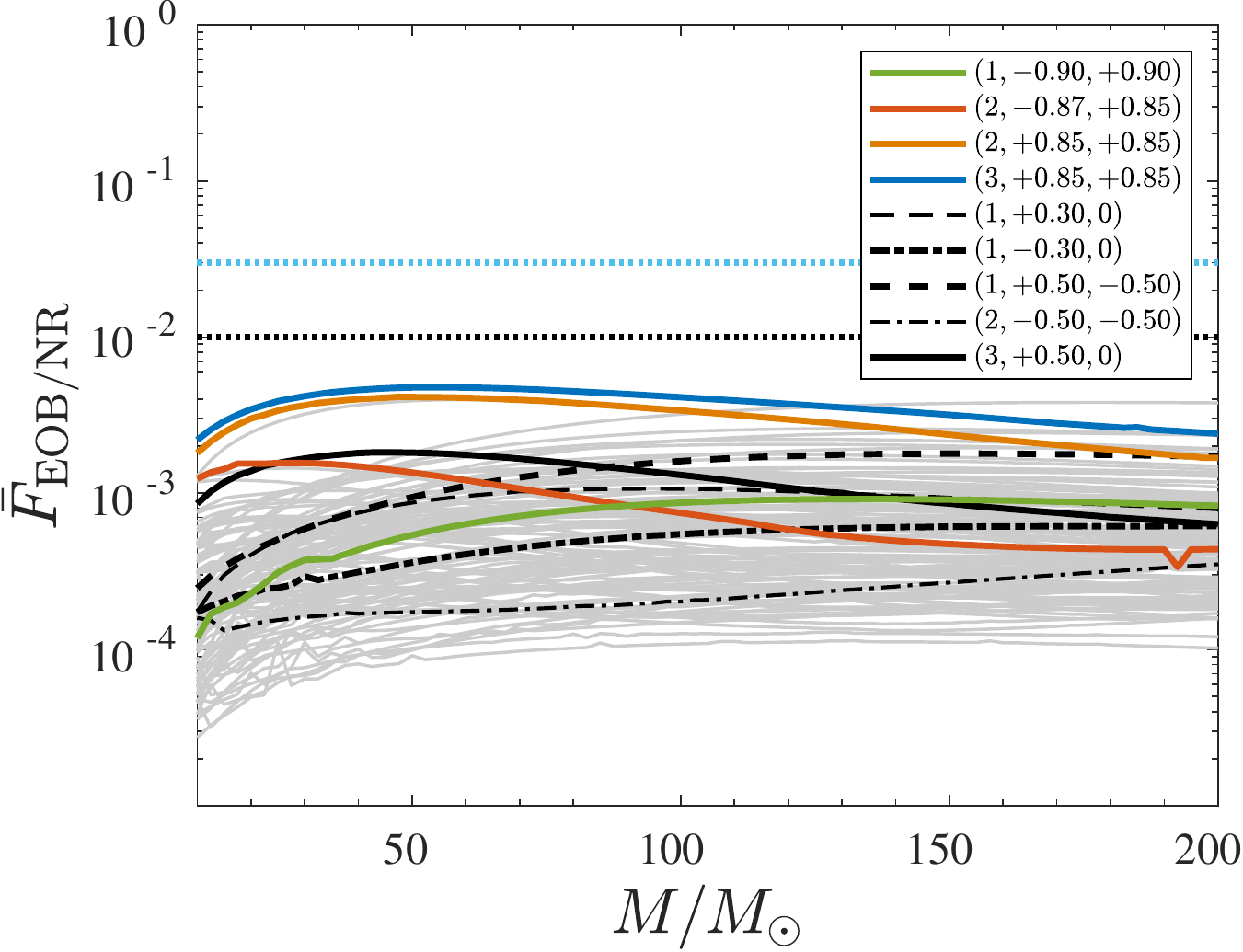}
    \includegraphics[width=0.32\textwidth]{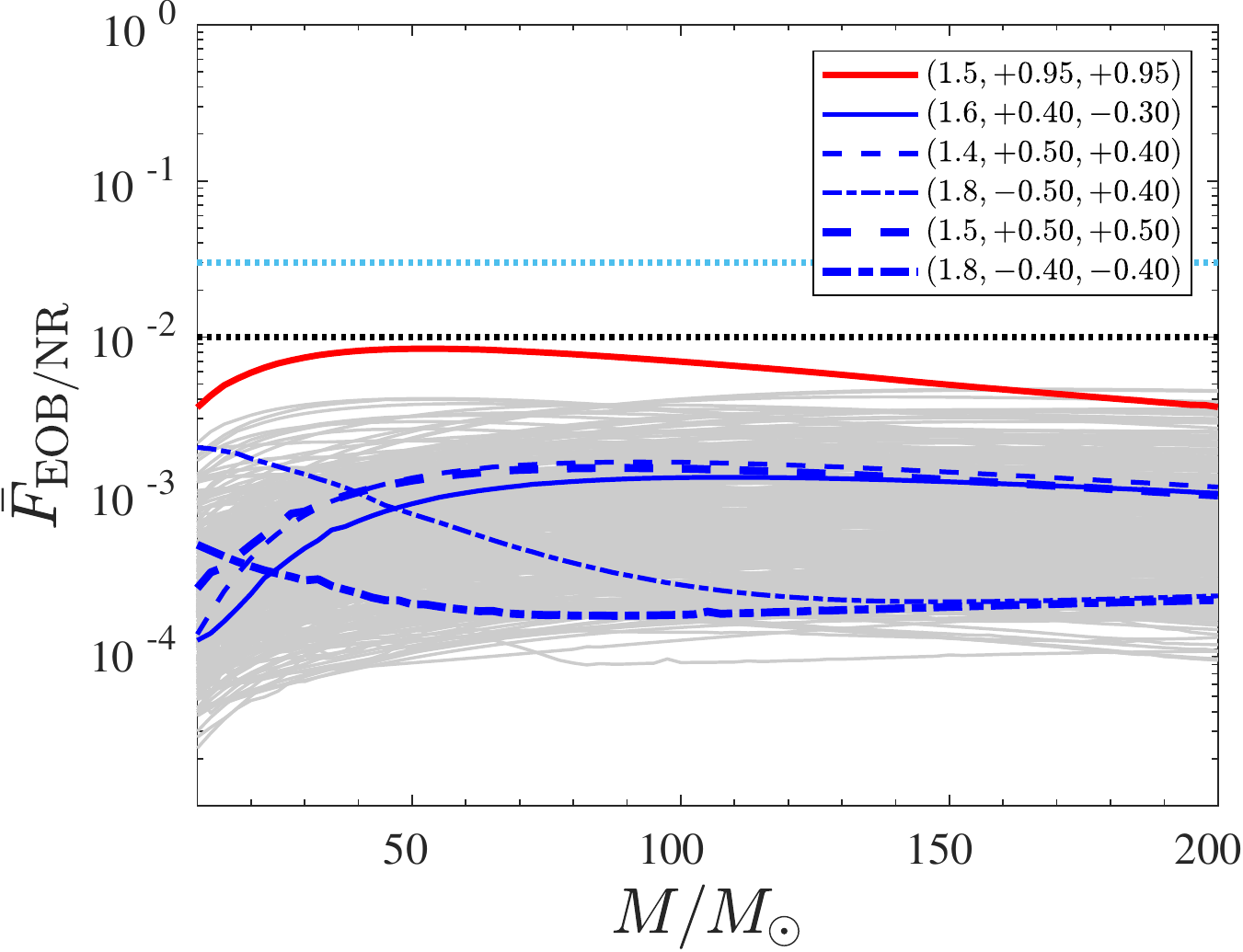}
    \includegraphics[width=0.32\textwidth]{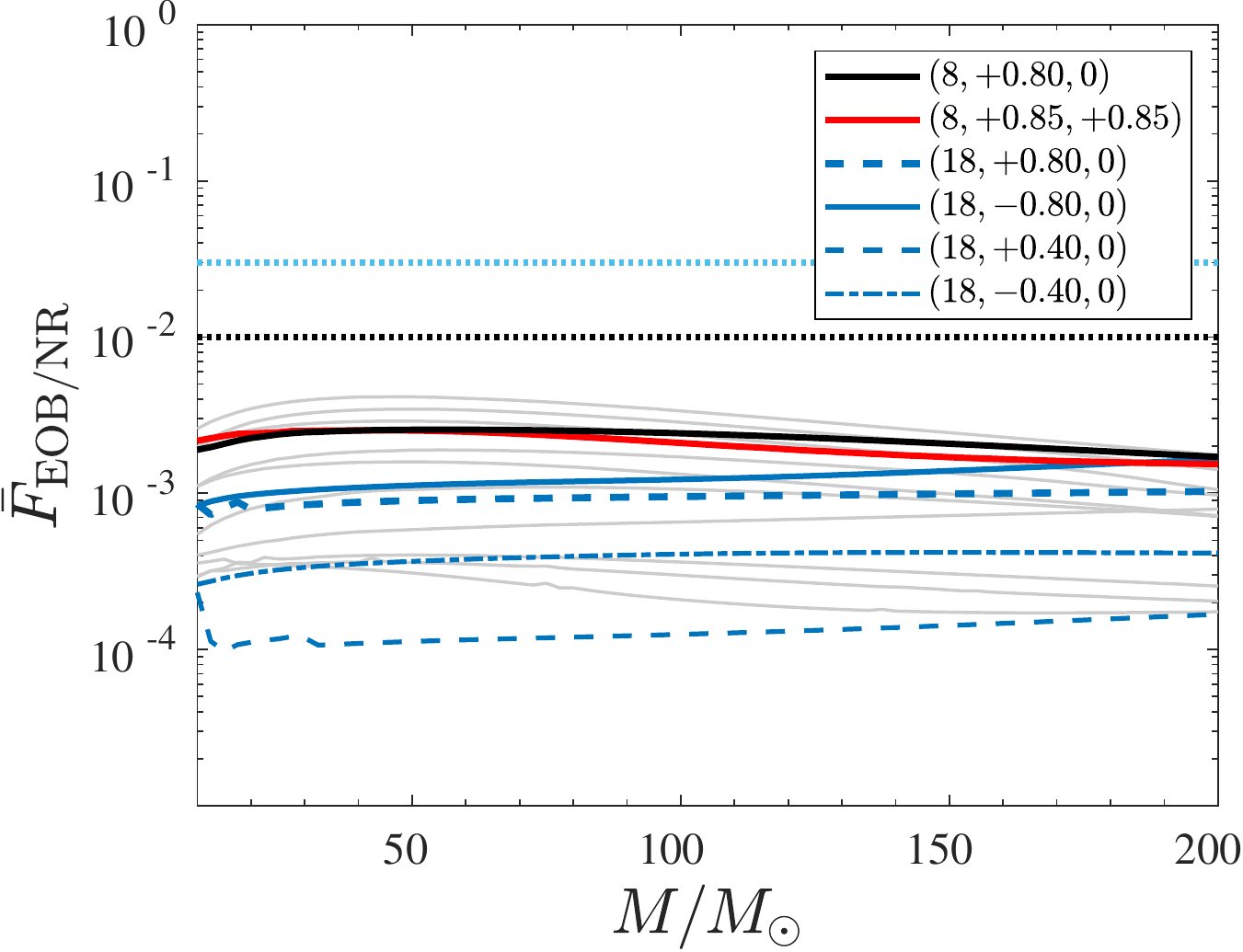}
    \caption{\label{fig:barF_iter}
    EOB/NR unfaithfulness for the $\ell=m=2$ mode using all currently available SXS
    NR simulations (left and middle panel) and a selection of BAM simulations (right panel).
    as published in Ref.~\cite{Nagar:2020pcj} (Paper~I). The NQC
    parameters are determined iteratively and included in the
    radiation reaction. The Hamilton equation of motion are solved numerically without
    using the post-adibatic approximation.
    Left panel: computation using SXS waveforms publicly released before February 3, 2019. 
    Middle panel: using SXS waveform data publicly released after February 3, 2019.
    $\bar{F}^{\rm max}_{\rm EOB/NR}$ is always below $0.4\%$ except for a single outlier, 
    red online, that however never exceeds $0.85\%$. The plot includes five exceptionally long 
    waveforms, each one developing more than 139 GW cycles before 
    merger,  SXS:BBH:1412, 1413, 1414, 1415 and 1416 (blue online).
    Right panel: same computation done with  a few BAM waveform data
    that include configurations with mass ratio $q=18$. See Ref.~\cite{Nagar:2020pcj} 
    for further details.}
  \end{center}
\end{figure*}

This appendix reports for completeness the faithfulness published in
Ref.~\cite{Nagar:2020pcj} (Paper~I) and obtained with the iterative NQC procedure and the full
ODE integration. The plots are shown in Fig.~\ref{fig:barF_iter} and can be directly 
compared to those shown in Fig.~\ref{fig:barF} in the main text, that are instead 
obtained with the fits for the NQC parameters $(a_1^{22},a_2^{22})$ entering the 
radiation reaction and the PA approximation to the numerical solution of the EOB Hamilton's equations 
during the inspiral. The unfaithfulness plots are obtained using the most recent realization
of the zero-detuned, high-power noise spectral density of Advanced LIGO~\cite{aLIGODesign_PSD}.

\section{NQC fits of $(a^{(2,2)}_1,a^{(2,2)}_2)$}
\label{sec:fit}

This appendix summarizes the NQC fits performed in this work.
The fits are performed hierarchically in different sectors of the parameter.
All fits have been performed with {\tt fitnlm} of {\tt Matlab}. The
superscript $(2,2)$ is dropped in the notation in this appendix.

\subsection{Non-spinning sector}
The fits in the non-spinning sector are obtained with a total of $27$ waveforms,
for mass-ratios $1\leq q\leq 30$.
The coefficient $a_1$ is fitted against $X_{12}^2=\left(1-4\nu\right)^2$ with the template
\begin{align}
	a_1\;\;\;\;	& = \frac{a_1^{q=1}}{1+b^{a_1}_1X_{12}^2+b^{a_1}_2X_{12}^4}
\end{align}
with 
\begin{align}
	a_1^{q=1} 	& = \;\;\,0.070974\nonumber\\
	b^{a_1}_1\,\;\; & = \;\;\,0.786350\nonumber\\
	b^{a_1}_2\,\;\; & = -9.085105 \ .\nonumber
\end{align}
The value of $a_1^{q=1}$ is extracted from $q=1$ NR data.

The coefficient $a_2$ in the non-spinning sector is fitted against $X_{12}=\sqrt{1-4\nu}$ with the template
\begin{align}
a_2\;\;\;\;	& = a_2^{q=1}\frac{1+b^{a_2}_1X_{12}+b^{a_2}_2X_{12}^2}{1+b^{a_2}_3X_{12}}
\end{align}
with
\begin{align}
a_2^{q=1} 		& = \;\;\,1.315133\nonumber\\
b^{a_2}_1\,\;\; & = -0.324849\nonumber\\
b^{a_2}_2\,\;\; & = -0.304506\nonumber\\
b^{a_2}_3\,\;\; & = -0.371614 \ .\nonumber
\end{align}
The value of $a_2^{q=1}$ is extracted from $q=1$ NR data.

\subsection{Equal-mass sector} 
Equal-mass data are defined by $\nu>0.2485$. A total of $40$ waveforms with spins $-0.98\leq \chi_{1,2}\leq 0.99$ are used to obtain the fits of the equal-mass region.
The coefficient $a_1$ in the equal-mass cases is fitted with the template:
\begin{align}
  a_1 	= c^{a_1}_0\frac{1+c^{a_1}_1\hat{S} + c^{a_1}_2\hat{S}^2 + c^{a_1}_3\hat{S}^3 + c^{a_1}_4\hat{S}^4}{
    1+c^{a_1}_5\hat{S} + c^{a_1}_6\hat{S}^2 + c^{a_1}_7\hat{S}^3}\ .
\end{align}
with the coefficients:
\begin{align}
c^{a_1}_0& =\,\; 0.121187 &\;\;\; c^{a_1}_1& =\,\; -5.950663 \nonumber\\	
c^{a_1}_2& =\,\; 9.420324 &\;\;\;c^{a_1}_3& = -10.601339\nonumber\\
c^{a_1}_4& = 17.641549 &\;\;\;c^{a_1}_5& =\,\; -5.684777 \nonumber\\
c^{a_1}_6& = 10.910451 &\;\;\;c^{a_1}_7& =\,\; -6.867377 \ . \nonumber
\end{align}

The coefficient $a_2$ is fitted to the same template. The fitted coefficients are:
\begin{align}
c^{a_2}_0& = \;\;\,1.331703 &\;\;\;c^{a_2}_1& = -4.237724 \nonumber\\	
c^{a_2}_2& = \;\;\,1.786023 &\;\;\;c^{a_2}_3& = \,\,10.546205\nonumber\\
c^{a_2}_4& = -9.698233 &\;\;\;c^{a_2}_5& = -6.225823 \nonumber\\
c^{a_2}_6& = \,\,13.209381 &\;\;\;c^{a_2}_7& = -9.402513 \ . \nonumber
\end{align}

\subsection{Sector with mass ratio $1<q<4$}
In this sector the fit of $a_1$ differs in two ways from the previous:
(i) the fit is factorized in a spinning part $a_1^S$ and a non-spinning part $a_1^0$,
and (ii) the fit uses the spin variable $\hat{S}_\nu\equiv \hat{S}/(1-2\nu)$. 
The full template is:
\begin{align}
a_1 & = a_1^0\cdot a_1^S\ ,\\	
a^0_1 & = d^{a_1}_0 \frac{1+ d^{a_1}_1\nu + d^{a_1}_2\nu^3}{1+d^{a_1}_3\nu}\ ,\\
a^S_1 & = \frac{1+d^{a_1}_4\hat{S}_n + d^{a_1}_5\hat{S}_\nu^2 + d^{a_1}_6\hat{S}_\nu^3 + d^{a_1}_7\hat{S}_\nu^4}{
	1+d^{a_1}_8\hat{S}_\nu + d^{a_1}_9\hat{S}_\nu^2 + d^{a_1}_{10}\hat{S}_\nu^3}\ .
\end{align}
The fitted coefficients take the values of $a_1^0$ are:
\begin{align}
d^{a_1}_0& = \,\;0.26132647 &\;\;\;d^{a_1}_1& = -4.90302367 \nonumber\\
d^{a_1}_2& = 20.67036124 &\;\;\;d^{a_1}_3& = -3.17109808\ . \nonumber
\end{align}
Note these coefficients are fitted to waveforms for which $\chi_2=\pm 0.01$
and $\chi_1$ is chosen such that $\hat{S}_\nu=0$. This approach is taken
also for all of the following non-spinning factor fits. In total 70 waveforms with 
$\hat{S}_n=0$ and further $454$ with spin $-0.9<\chi_{1,2}\leq 0.99$. Of these $160$ are 
focused on the high positive region, $0.8\leq \chi_{1,2} \leq 0.99$.

The fitted coefficients of $a_1^S$ are:
\begin{align}
d^{a_1}_4& = -3.082861 &\;\;\;d^{a_1}_5& = 2.169948\nonumber\\
d^{a_1}_6& = -0.636353 &\;\;\;d^{a_1}_7& = 0.741419\nonumber\\
d^{a_1}_8& = -2.843896 &\;\;\;d^{a_1}_9& = 2.709697\nonumber\\
d^{a_1}_{10}& = -0.832894 & \ .&\nonumber
\end{align}

The coefficient $a_2$ is fitted in a factorized form as well. Additionally, it holds an explicit 
dependency of $a_2^S$ on $\nu$:
\begin{align}
a_2 & = a_2^0\cdot a_2^S\ ,\\	
a^0_2 & = d^{a_2}_0 \frac{1+ d^{a_2}_1\nu + d^{a_2}_2\nu^3}{1+d^{a_2}_3\nu}\ ,\\
a^S_2 & = \frac{1+d^{a_2}_4\hat{S}_\nu + d^{a_2}_5\hat{S}_\nu^2 + d^{a_2}_6\hat{S}_\nu^3 + d^{a_2}_7\hat{S}_\nu^4}{
	1+d^{a_2}_8\hat{S}_\nu + d^{a_2}_9\hat{S}_\nu^2 + d^{a_2}_{10}\hat{S}_\nu^3}\ ,\\
d^{a_2}_i & = d^{a_2}_{i,0} \left(1+d^{a_2}_{i,1}\nu\right), \ for\ i=4,...,10\ .
\end{align}
The fitted coefficients of $a_2^0$ are:
\begin{align}
d^{a_2}_0& = \;\;\, 1.03364144 &\;\;\;d^{a_2}_1& = -3.46191440 \nonumber\\
d^{a_2}_2& = -7.86652243 &\;\;\;d^{a_2}_3& = -3.96268815\ . \nonumber
\end{align}
The fitted coefficients of $a_2^S$ are:
\begin{align}
d^{a_2}_{4,0}& =\;\;\, 0.036452 &\;\;\;d^{a_2}_{4,1}& = -64.360789 \nonumber\\
d^{a_2}_{5,0}& =\;\;\, 0.275707 &\;\;\;d^{a_2}_{5,1}& = -34.573145 \nonumber\\
d^{a_2}_{6,0}& = -0.113951 &\;\;\;d^{a_2}_{6,1}& =\;\;\,\,\; 0 \nonumber\\
d^{a_2}_{7,0}& = -2.531304 &\;\;\;d^{a_2}_{7,1}& =\,\; -7.691661 \nonumber\\
d^{a_2}_{8,0}& = -1.025824 &\;\;\;d^{a_2}_{8,1}& = \;\;\,\,\;4.237539 \nonumber\\
d^{a_2}_{9,0}& =\;\;\,  0.593579&\;\;\;d^{a_2}_{9,1}& =\;\;\,\,\; 1.661809 \nonumber\\
d^{a_2}_{10,0}& = -0.939736 &\;\;\;d^{a_2}_{10,1}& = \,\;-6.333442\ . \nonumber
\end{align}
$d^{a_2}_{6,1}$ is set to $0$ prior to the evaluation of the fit to improve 
the convergence of the fit.

\subsection{Sector with mass ratio $q\geq 4$}
For the following fits a similar approach to was taken as above.
A total of $44$ with $\hat{S}_n=0$ have been generated. 
$186$ waveforms with $-0.99\leq\chi_{1,2}\leq 0.99$ have been used to capture the $q=4$ behavior accurately.
$1470$ further waveforms with $-0.99\leq\chi_{1,2}\leq 0.85$ have been used 
to fit the extrapolation of the $q=4$ fit up to mass ratio $q=30$.
The coefficient $a_1$ for $q\geq 4$ has an additional feature. The explicit $\nu$ 
dependence is fitted through $x_{\nu}=\nu-0.16$. The full template is:
\begin{align}
a_1 & = a_1^0\cdot a_1^S\ ,\\	
a^0_1 & = e^{a_1}_0 \frac{1+ e^{a_1}_1\nu + e^{a_1}_2\nu^3}{1+e^{a_1}_3\nu}\ ,\\
a^S_1 & = \frac{1+e^{a_1}_4\hat{S}_\nu + e^{a_1}_5\hat{S}_\nu^2 + e^{a_1}_6\hat{S}_\nu^3 + e^{a_1}_7\hat{S}_\nu^4}{
	1+e^{a_1}_8\hat{S}_\nu + e^{a_1}_9\hat{S}_\nu^2 + e^{a_1}_{10}\hat{S}_\nu^3}\ ,\\
e^{a_1}_i & = e^{a_1}_{i,0}\frac{1+e^{a_1}_{i,1}x_\nu}{1+e^{a_1}_{i,2}x_\nu} , \ for\ i=4,...,10\ .
\end{align}
The fitted $a_1^0$ coefficients are:
\begin{align}
e^{a_1}_0& = \;\;\,0.341803 &\;\;\;e^{a_1}_1& = -1.350488 \nonumber\\
e^{a_1}_2& = -6.353357 &\;\;\;e^{a_1}_3& = \;\;\,2.216156\ . \nonumber
\end{align}
The coefficients of $a_1^S$ are fitted in 2 steps. First,
for $q=4$ and second, an extrapolated fit from there. 
The coefficients $e^{a_1}_{i,0}$ are fitted to $q=4$:
\begin{align}
e^{a_1}_{4,0}& = -2.287721 &\;\;\;e^{a_1}_{5,0}& = -0.598451 \nonumber\\
e^{a_1}_{6,0}& = \;\;\, 0.766069 &\;\;\;e^{a_1}_{7,0}& = \;\;\, 1.857169\nonumber\\
e^{a_1}_{8,0}& = -2.035234 &\;\;\;e^{a_1}_{9,0}& = \;\;\, 0.836427\nonumber\\
e^{a_1}_{10,0}& =\;\;\, 0.297476 & \ . &\nonumber
\end{align}
The remaining coefficients model the extrapolation of the spin dependence to larger mass ratios and are:
\begin{align}
e^{a_1}_{4,1}& = \;\;\,\;\,7.650946 &\;\;\;e^{a_1}_{4,2}& =\;\;\,\;\,7.106992 \nonumber\\
e^{a_1}_{5,1}& = -60.630748&\;\;\;e^{a_1}_{5,2}& = -69.630357\nonumber\\
e^{a_1}_{6,1}& = \;\;\,47.114247 &\;\;\;e^{a_1}_{6,2}& =\;\;\, \;\,5.733002\nonumber\\
e^{a_1}_{7,1}& = -12.905707 &\;\;\;e^{a_1}_{7,2}& =\;\;\,\;\, 5.045688\nonumber\\
e^{a_1}_{8,1}& = \;\;\,\;\,3.515869&\;\;\;e^{a_1}_{8,2}& = \;\;\,\;\,1.564146\nonumber\\
e^{a_1}_{9,1}& = \;\;\,\;\,0.642864&\;\;\;e^{a_1}_{9,2}& =\;\;\,\;\, 2.947890 \nonumber\\
e^{a_1}_{10,1}& = \;\;\,31.023038 &\;\;\;e^{a_1}_{10,2}& =\;\;\,\;\, 1.829543 \ .\nonumber
\end{align}

The coefficient $a_2$ is fitted similarly with the template:
\begin{align}
a_2 & = a_2^0\cdot a_2^S\ ,\\	
a^0_2 & = e^{a_2}_0 \frac{1+ e^{a_2}_1\nu + e^{a_2}_2\nu^3}{1+e^{a_2}_3\nu}\ ,\\
a^S_2 & = \frac{1+e^{a_2}_4\hat{S}_\nu + e^{a_2}_5\hat{S}_\nu^2 + e^{a_2}_6\hat{S}_\nu^3 + e^{a_2}_7\hat{S}_\nu^4}{
	1+e^{a_2}_8\hat{S}_\nu + e^{a_2}_9\hat{S}_\nu^2}\ ,\\
e^{a_2}_i & = e^{a_2}_{i,0}\frac{1+e^{a_2}_{i,1}x_\nu}{1+e^{a_2}_{i,2}x_\nu} , \ for\ i=4,...,9\ .
\end{align}
The fitted $a_2^0$ coefficients are:
\begin{align}
e^{a_2}_0& =  \;\,\;\;\, 0.929192 &\;\;\;e^{a_2}_1& =\;\;\, 1.334263 \nonumber\\
e^{a_2}_2& =  -26.389790&\;\;\;e^{a_2}_3& = -1.289984\ . \nonumber
\end{align}
The coefficients of $a_2^S$ are fitted in 2 steps as well. 
The coefficients $e^{a_2}_{i,0}$ have been fitted to $q=4$:
\begin{align}
e^{a_2}_{4,0}& = -0.886561 &\;\;\;e^{a_2}_{5,0}& = -1.953955 \nonumber\\
e^{a_2}_{6,0}& =\;\;\, 1.366537 &\;\;\;e^{a_2}_{7,0}& = \;\;\,0.950212 \nonumber\\
e^{a_2}_{8,0}& = -2.531000 &\;\;\;e^{a_2}_{9,0}& =\;\;\, 1.723991\ .\nonumber
\end{align}

The remaining coefficients model the extrapolation of the spin dependence to larger mass ratios and are:
\begin{align}
e^{a_2}_{4,1}& =\;\;\, 15.871482 &\;\;\;e^{a_2}_{4,2}& =\;\,\;\;\, 5.066190 \nonumber\\
e^{a_2}_{5,1}& =\;\,\;\;\, 7.168498 &\;\;\;e^{a_2}_{5,2}& =\;\,\;\;\, 6.709490 \nonumber\\
e^{a_2}_{6,1}& =\;\;\, 18.583382 &\;\;\;e^{a_2}_{6,2}& =\;\,\;\;\, 5.764512 \nonumber\\
e^{a_2}_{7,1}& = -14.038564 &\;\;\;e^{a_2}_{7,2}& = -17.126231 \nonumber\\
e^{a_2}_{8,1}& =\;\,\;\;\, 6.387917 &\;\;\;e^{a_2}_{8,2}& =\;\,\;\;\, 3.438456 \nonumber\\
e^{a_2}_{9,1}& =\;\,\;\;\, 8.867098 &\;\;\;e^{a_2}_{9,2}& =\;\,\;\;\, 2.910938 \ . \nonumber
\end{align}

\bibliography{refs20210417.bib,local.bib}

\end{document}